\documentclass[twocolumn,aps,pra,showpacs,eqsecnum,%
floatfix]{revtex4}

\usepackage{dcolumn}
\usepackage{graphicx}
\usepackage{epsf}
\usepackage{amsmath}
\usepackage{bm}
\usepackage{dcolumn}
\usepackage{nicefrac}
\usepackage{times}

\newcolumntype{.}{D{x}{}{-1}}

\newcommand{\addrHD}{Max--Planck--Institut f\"ur Kernphysik,
Saupfercheckweg 1, 69117 Heidelberg, Germany}

\newcommand{\addrFR}{Physikalisches Institut der
Albert--Ludwigs--Universit\"{a}t, Theoretische Quantendynamik,\\
Hermann--Herder--Stra\ss{}e 3, 79104 Freiburg im Breisgau, Germany}

\begin{document}

\bibliographystyle{myprsty}

\title{Relativistic and Radiative Corrections to the Mollow Spectrum}

\author{J\"org \surname{Evers}}
\email{evers@mpi-hd.mpg.de}

\author{Ulrich D.~\surname{Jentschura}}
\email{ulj@mpi-hd.mpg.de}

\author{Christoph H. \surname{Keitel}}
\email{keitel@mpi-hd.mpg.de}

\affiliation{\addrHD}
\affiliation{\addrFR}

\begin{abstract}
The incoherent, inelastic part of the resonance fluorescence spectrum
of a laser-driven atom is known as the Mollow spectrum
[B. R. Mollow, Phys. Rev. {\bf 188}, 1969 (1969)].
Starting from this level of description, we discuss theoretical 
foundations of high-precision spectroscopy using the resonance fluorescence 
light of strongly laser-driven atoms. Specifically, we evaluate the leading
relativistic and radiative corrections to the 
Mollow spectrum, up to the relative orders of $(Z\alpha)^2$
and $\alpha\,(Z\alpha)^2$, respectively, and Bloch--Siegert shifts
as well as stimulated radiative corrections involving off-resonant virtual
states. Complete results are provided for
the hydrogen 1$S$-2$P_{\nicefrac{1}{2}}$ and
1$S$-2$P_{\nicefrac{3}{2}}$ transitions; these include all relevant 
correction terms up to the specified order of approximation
and could directly be compared to
experimental data. As an application, the outcome of such experiments 
would allow for a sensitive test of the validity of the dressed-state basis
as the natural description of the combined atom-laser system.

\end{abstract}


\maketitle

\section{\label{intro}INTRODUCTION}

Experimental possibilities in high-precision spectroscopy have received a
rather significant `boost' in recent years due to the availability of
phase coherent regularly spaced frequency combs that may bridge large
frequency intervals between frequency standards and optical transition
frequencies~\cite{ReEtAl2000}. In general terms, highly accurate
spectroscopy may lead to an experimental verification of known theoretical
models of the physical process under study. Precision measurements ---in
combination with theory--- allow to obtain accurate values for physical
parameters or fundamental constants~\cite{MoTa2000}. With increasing
accuracy, one may even ask whether the so-called constants are in fact
constant~\cite{PrTjMa1995,
WeFlChDrBa1999,DzFlWe1999,WeEtAl2003,MuWeFl2003,%
Uz2003,MaEtAl2003,FiEtAl2004}.  These results may as well be used as input
to more applied physics as the creation of unit standards e.g. for time
and mass.  

On the theoretical side, quantum electrodynamics is one of the most
accurate theories known so far. In many previous studies, the $S$-matrix
formalism has been used to obtain accurate predictions for experimental
investigations. The $S$-matrix relates the distant past to the distant
future without referring to the dynamics of the intermediate times and
leads effectively to a somewhat static description~\cite{Pa1991}. 

Thus, the $S$-matrix formalism cannot, {\em a priori}, lead to a satisfactory
description of quantum electrodynamic corrections to dynamical
processes, and the subject of this paper 
is to provide a first step in the direction of a high-precision
theoretical description of dynamical processes including radiative
corrections, using a laser-driven atom as a paradigmatic 
example. Obviously, the treatment of radiative corrections 
to a dynamically driven atomic transition requires
input from two different areas, which are laser physics
and quantum electrodynamics. While the two areas are related,
there are a couple of subtle points to consider when a 
unified understanding of a specific problem is sought,
whose nature inevitably requires concepts introduced within
the context of either of the two areas. In particular,
it is known that the description
of dynamical processes requires considerable care in 
the treatment of the gauge-dependence of amplitudes,
and with regard to the physical interpretation of 
the wave functions used in the mathematical 
description~\cite{LaRe1950,La1952,PoZi1959,Ya1976,FoQuBa1977,Ko1978prl,%
BrScScZuGo1983,BeScSc1984,ScBeBeSc1984,LaScSc1987}.

A classic textbook example for a dynamical atom-laser system, well-known in 
theoretical quantum optics~\cite{ScZu1997}, consists of the
Jaynes--Cummings model of an atom that contains two relevant energy
levels interacting with a single monochromatic laser field
mode~\cite{JaCu1963}. Due to the driving of the laser field, the atomic
population undergoes Rabi oscillations. The population is driven
periodically from the upper to the lower state and vice versa. The
emission spectrum of this process with a strong driving field is known as
the Mollow spectrum~\cite{Mo1969}. This case of strong driving may easily
be interpreted in terms of the so-called dressed states. Laser-dressed
states are defined
as the eigenstates of the combined system of atom and driving laser
field~\cite{CT1975misc} and have proven to be useful in countless cases of
both theory and experiment, one of which is the Autler-Townes
splitting~\cite{AuTo1955}.

\begin{figure}[b]
\includegraphics[width=6cm]{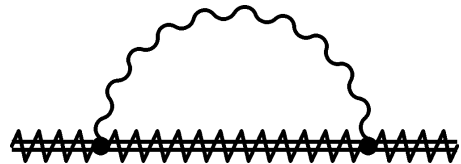}
\caption{\label{fig1}
Diagrammatic representation of a radiative self-energy correction to the
laser-dressed atomic state. The double line corresponds to the electron
bound by the nuclear Coulomb field. 
The jagged line denotes the additional dressing of the
bound electron by the (strong) laser field. 
The self-energy of a laser-dressed, Coulomb-bound 
electron is a quantum-field theoretic problem in the presence 
of two classical background fields.}
\end{figure}

When evaluating radiative corrections to the Mollow spectrum,
it is natural to start from the dressed-state basis, which consists of
the natural eigenstates of the (strongly) coupled atom-laser system rather than 
the bare atomic states. It might be assumed that in order to fully treat the 
Lamb shift of laser-dressed states, it would be sufficient to 
simply correct the energies of the bare states that enter
into the formulas for the generalized Rabi frequencies
by the ``bare-state'' (i.e.,~the usual, ordinary) Lamb shift.
Indeed, the first investigations on the 
problem~\cite{Kr1982} revealed corrections to the 
dressed-state ``quasi-energies'' consistent with this 
assumption.
However, recently, it was found that at nonvanishing detuning and 
Rabi frequency, the Lamb shift of dressed states is 
nontrivially different from the bare Lamb shift~\cite{JeEvHaKe2003,JeKe2004}.
Thus the distinction between evaluations in terms of the bare-
and the dressed state basis in fact has to be made.
In the limit of vanishing detuning, the coincidence of
the bare- and dressed Lamb shift effects on the detuning is obtained
only after a summation of a specific series whose leading correction
term may be obtained by carrying out the calculation to
second order in the atom-field interaction.
A diagrammatic representation of a
radiative correction to the dressed state is shown in
Fig.~\ref{fig1}. 

Thus in this article we present a detailed and complete theoretical
analysis of the leading nonrelativistic and relativistic corrections
to the Mollow spectrum, up to the relative orders of $(Z\alpha)^2$
and $\alpha\,(Z\alpha)^2$, respectively, and of Bloch--Siegert shifts
as well as stimulated radiative corrections involving off-resonant virtual
states, and laser-field configuration dependent corrections. 
The purpose is to enable a direct comparison between theory 
and experiment for a high-precision spectroscopic
investigation involving laser-dressed states. 
Such a comparison to experimental data would allow to
address questions related to the physical reality of the dressed states 
(and their ``quasi-energy'') on
the one hand, and of the nature and the interpretation of the various
radiative corrections on the other hand.  As a promising candidate for the
experiment, we identify the resonance fluorescence spectrum of a strongly
driven hydrogen 1$S$-2$P$ transition, which to lowest order may be
described by the standard Mollow spectrum. 
A coherent Lyman-$\alpha$ source~\cite{EiWaHa2001,Pa2002} 
has recently become available as a driving field, and we
show that ionization into the continuum does not prohibit an experimental
implementation. In particular, we discuss corrections which are due to 
resonant and off-resonant excitations as well as the Bloch-Siegert shift, 
and corrections to the transition dipole moment and to the secular
approximation leading to the Mollow spectrum. As a result, we provide
theoretical predictions which are directly comparable to possible
experimental data.

The article is organized as follows. In Sec.~\ref{results} we
introduce our system of interest and provide the relevant theoretical
background for the further analysis. In Sec.~\ref{calculation}, we evaluate
the corrections to the Mollow spectrum, which we divide into modifications
of the detuning (Sec.~\ref{sCorrDetuning}) 
and the Rabi frequency (Sec.~\ref{sCorrRabi}).
The dominant 
relativistic corrections are of the order of $(Z\alpha)^2$, where
$Z$ is the nuclear charge number, and leading radiative effects 
lead to correction terms of the order of $\alpha\,(Z\alpha)^2\,
\ln[(Z\alpha)^{-2}]$.
In Sec.~\ref{num-analysis}, we provide numerical data for the
hydrogen $1S$--$2P_j$ transitions ($j=\nicefrac{1}{2},\nicefrac{3}{2}$).
Sec.~\ref{discuss} discusses and  summarizes the results.

\section{\label{results}MOLLOW SPECTRUM}

In this section, we introduce our system and recall results of previous
studies which will serve as the basis of our analysis. Throughout the
calculations, we adopt natural units with $\hbar = \epsilon_0 = c = 1$.
The electron mass is denoted by $m$. We make use of the Einstein summation
convention unless stated otherwise, and we employ the
length gauge for all wave functions and 
operators as we deal with off-resonant excitations,
as it is done in most of the literature, 
and in textbooks on the subject~(see e.g.~\cite{ScZu1997}). 
The Mollow spectrum contains the incoherent, inelastic
part of the atomic fluorescence, i.e.~the fluorescence 
spectrum mediated by the many-photon processes whose
intensity dominates over the elastic part in a strongly
driven atom-laser system. In a purely quantum electrodynamic
formalism, the description of many-photon processes would 
require perturbation theory in exceedingly high orders.
However, as is well known, the description using 
dressed states~\cite{CT1975misc} allows for a considerable 
simplification, as the formulas for the Mollow spectrum~\cite{Mo1969}
follow rather naturally in terms of transitions
among the dressed atomic states which incorporate the 
atom-laser interaction to all orders in perturbation theory.

Before we now start with the discussion of the Mollow
spectrum, a slight detour on questions related to 
gauge transformations of the laser-atom interaction is in order.
The ``length gauge'' means that the laser-atom interaction
is formulated in terms of the interaction $-q \,
\bm{E_{\rm L}}\cdot\bm{x}$, where $q$ is the physical electron
charge and $\bm{E_{\rm L}}$ is the 
(gauge-invariant, observable) laser field strength. Instead, in 
the ``velocity gauge'', the interaction
is formulated as $-q \, \bm{A_{\rm L}}\cdot\bm{p}/m$,
where $\bm{A_{\rm L}}$ is a gauge-dependent,
suitable vector potential for the 
laser field. In the velocity gauge, of course, one also 
has to add the $\bm{A_{\rm L}}^2$ term, but dipole interactions
are mediated exclusively by the ``$\bm{A_{\rm L}}\cdot\bm{p}$''-interaction.
Due to gauge invariance, the two possible gauges are equivalent,
{\em provided} that the gauge transformation of the 
wave function is properly taken into account~\cite{ScZu1997,CTDRGr1992}. 
In order to avoid confusion, we stress here the
absolute necessity of of considering the gauge transformation
of the wave function in dynamical 
processes~\cite{LaRe1950,La1952,PoZi1959,Ya1976,FoQuBa1977,Ko1978prl,%
BrScScZuGo1983,BeScSc1984,ScBeBeSc1984,LaScSc1987}. 
According to~\cite{La1952,LaScSc1987}, the usual physical interpretation
of a Schr\"{o}dinger wave function is only conserved if the 
length gauge is used for the description of the atom-laser interaction.
For dynamical processes, the velocity gauge leads to many
more terms in intermediate steps of the calculation than the 
length-gauge formulation chosen here, due to the necessity 
of applying the gauge transformation to the wave function.
Nevertheless, we would like to mention the possibility 
of an independent verification of our derivation, as presented here,
in different gauges. In this case, the gauge transformation
of the wave function should be applied already on the level
of quantum mechanics
(i.e., on the level of the Mollow spectrum as discussed in the current 
Section), not just to the quantum electrodynamic corrections 
discussed in the following Sections. 
This concludes our detour regarding gauge transformations.

We recall from~\cite{Mo1969,JeEvHaKe2003,JeKe2004} that the incoherent resonance
fluorescence spectrum of atoms driven by a monochromatic coherent laser
field may be expressed as
\begin{eqnarray}
\label{MollowSpectrumExact}
{\mathcal S}_{\rm inc}(\omega) &=&
\frac{\Gamma}{\pi}\,
\frac{2 \Gamma^2 + \Omega^2 + 2 (\omega-\omega_{\rm L})^2}
{\Gamma^2 + 2\,\Omega^2 + 4\,\Delta^2}\,
\nonumber\\[1ex]
& & \times \frac{4\,\Gamma\,\Omega^4}
{X_0 + X_2 \, \Gamma^2 + X_4 \, \Gamma^4 + X_6 \, \Gamma^6}\, ,
\end{eqnarray}
where
\begin{subequations}
\begin{eqnarray}
X_0 &=& 16\,\left[\Delta^2 + \Omega^2 - (\omega-\omega_{\rm L})^2\right]^2\,
(\omega-\omega_{\rm L})^2\,, \\[2ex]
X_2 &=& 4\,\left[6\,(\omega-\omega_{\rm L})^4 -
2\,(3\,\Delta^2 - \Omega^2) \, (\omega-\omega_{\rm L})^2 \right.
\nonumber\\[1ex]
& & \left. + (2\,\Delta^2 + \Omega^2)^2\right]\,, \\[2ex]
X_4 &=& 8\,\Delta^2 + 4\,\Omega^2 + 9\,(\omega-\omega_{\rm L})^2\,,\\[2ex]
X_6 &=& 1\,.
\end{eqnarray}
\end{subequations}
Here, $\Omega$ is the Rabi frequency 
\begin{equation}
\Omega = -q \left <e|{\bm x}\cdot {\bm \epsilon}_{\rm L}|g \right >  
{\cal E}_{\rm L}
\end{equation}
of the driving laser field ${\bm E}_{\rm L}(t) = {\cal E}_{\rm L} {\bm
\epsilon}_{\rm L} \cos (\omega_{\rm L}t)$ with frequency $\omega_{\rm L}$,
macroscopic classical amplitude ${\cal E}_{\rm L}$ and polarization ${\bm \epsilon}_{\rm L}$. $q=-|q|$ is the electron charge.
$\Delta = \omega_{\rm L} - \omega_{eg}$ is the detuning of the laser field
frequency from the atomic transition frequency $\omega_{eg}$, and $\Gamma$
is the spontaneous decay rate of the atomic transition. The excited and
the ground state of the laser-driven transition are denoted by $|e\rangle$
and $|g\rangle$, respectively, and ${\bm x}$ is the position operator
vector.
In secular approximation $\Omega \gg \Gamma$, this expression simplifies to
\begin{eqnarray}
\label{MollowSpectrumSecular}
\lefteqn{{\cal S}_{\rm inc}(\omega) \approx
\frac{\Gamma}{\pi}\, \left[
\frac{\Gamma_0 \, A_0}
  {(\omega - \omega_{\rm L})^2 + \Gamma^2_0}
\right.}
\\[2ex]
& & \left. + \frac{\Gamma_+ \, A_+}
  {(\omega - \omega_{\rm L} - \Omega_{\rm R})^2 + \Gamma_+^2}
+ \frac{\Gamma_- \, A_-}
  {(\omega - \omega_{\rm L} + \Omega_{\rm R})^2 + \Gamma_-^2}
\right]\,.
\nonumber
\end{eqnarray}
Here, the separation of the Mollow spectrum into one
central peak located at $\omega=\omega_{\rm L}$
and into two sidebands shifted by the generalized Rabi
frequency $\Omega_{\rm R} = \sqrt{\Delta^2 + \Omega^2}$ may easily be
seen. The amplitudes and widths are given by
\begin{subequations}
\begin{eqnarray}
\label{Ainc0}
A_0 &=& \frac{\Omega^6}
  {4 \, \Omega_{\rm R}^2 \, (\Omega_{\rm R}^2 + \Delta^2)^2} \, ,
\\[2ex]
\label{Aplusminus}
A_\pm &=& \frac{\Omega^4}
  {8 \, \Omega_{\rm R}^2 \, (\Omega_{\rm R}^2 + \Delta^2)} \, ,
\\[2ex]
\label{Gamma0}
\Gamma_0 &=& \Gamma \, \frac{\Omega^2 + 2\,\Delta^2}{2 \Omega_{\rm R}^2} \,, 
\\[2ex]
\label{Gammaplusminus}
\Gamma_\pm &=&  
\Gamma \, \frac{3 \Omega^2 + 2\,\Delta^2}{4 \Omega_{\rm R}^2} \,.
\end{eqnarray}
\end{subequations}
The approximate 
form (\ref{MollowSpectrumSecular}) does not represent the
positions of the sideband peaks accurately in cases where
$\Gamma/\Omega_{\rm R}$ is not small. Indeed,
the position $\omega_\pm$ of the sideband peaks may be expanded 
in a series in powers of
$\Gamma/\Omega_{\rm R}$ whose first terms read
\begin{eqnarray}
\label{omegapm}
\lefteqn{\omega_\pm = \omega_{\rm L} \pm \Omega_{\rm R} \, \left[ 1 - 
\frac{4 + y^2}{8 \, (1 + y^2)} \,
\left(\frac{\Gamma}{\Omega_{\rm R}}\right)^2 \right.}
\nonumber\\[2ex]
& & \left. - \frac{70 + 8\,y^2 + y^4}{128\,(1+y^2)^2}\,
\left(\frac{\Gamma}{\Omega_{\rm R}}\right)^4 +
{\mathcal O}\left( \frac{\Gamma}{\Omega_{\rm R}}\right)^6 \right]\,, 
\label{secular-corrections}
\end{eqnarray}
with $y = \Delta/\Omega$.
For vanishing detuning $\Delta = 0$,
which implies $\Omega_{\rm R} = \Omega$,
 Eq.~(\ref{omegapm}) specializes to
\begin{eqnarray} 
\label{omegapmzerodelta}
\omega_\pm &=& \omega_{\rm L} \pm \Omega\, \left[ 1 -
\frac12\left(\frac{\Gamma}{\Omega}\right)^2 -
\frac{35}{64}\left(\frac{\Gamma}{\Omega}\right)^4 \right. \nonumber \\*
&& \qquad \qquad \left .  
+ {\mathcal O}\left( \frac{\Gamma}{\Omega}\right)^6 \right]\,.
\end{eqnarray}
The correction terms move the sideband peaks closer to the
central maximum.

The above results in secular approximation may easily be interpreted with the
help of the so-called dressed states, which are defined
as the eigenstates of the interaction part of the
Hamiltonian. Under the influence of the external driving 
field, the atomic states are no longer eigenstates of the
Hamiltonian, but rather have to be combined with the
driving laser field to give the new eigenstates.
To show the precise composition of the dressed states,
we use the quantum representations of the Rabi frequency 
\begin{equation}
\Omega_n = 2 \: g_{\rm L} \sqrt{n+1} \,,
\end{equation}
the mixing angle $\theta_n$ defined by
\begin{equation}
\label{defmixing}
\tan (2\theta_n) = -\Omega_n / \Delta\,,
\end{equation}
and the generalized Rabi frequency
\begin{equation}
\Omega_{\rm R}^{(n)} = \sqrt{\Omega_n^2 + \Delta^2}
\end{equation}
rather than the corresponding classical entities.
Here, $n$ is the number of photons in the laser field mode,
and the coupling constant $g_{\rm L}$ for the interaction
of the driving laser field with the main atomic transition
is defined by
\begin{equation}
g_{\rm L} = 
- q \, \langle g | \bm{\epsilon}_{\rm L} \cdot \bm{x} | e \rangle \,
{\cal E}_{\rm L}^{\rm (\gamma)}\, ,
\end{equation}
where ${\cal E}_{\rm L}^{\rm (\gamma)} = 
\sqrt{\omega_{\rm L}/2V}$ is the electric laser field per photon
and $V$ is the quantization volume. The matching of the electric
field per photon with the corresponding classical macroscopic
electric field ${\cal E}_{\rm L}$ is given by
\begin{equation}
\label{matching}
2 \: \sqrt{n+1} \: {\cal E}_{\rm L}^{\rm (\gamma)} \: 
\longleftrightarrow \: {\cal E}_{\rm L} \,.
\end{equation}
Throughout this article, we will sometimes refer to
the quantum description during the derivations, but use
the classical entities in the final results. We may
switch between the two descriptions as the driving
laser field is assumed to be intense in our analysis.
The matching of the quantum and the classical entities
is possible with the help of the following list of replacements:
\begin{subequations}
\label{matchOmega}
\begin{eqnarray}
\Omega_n &\leftrightarrow& \Omega \, ,\\
\label{matchmixing}
\theta_n &\leftrightarrow& \theta \, ,\\
n+1 &\approx& n \, .
\end{eqnarray}
\end{subequations}
Using this notation, the dressed states are given by
\begin{subequations}
\label{dressedstates}
\begin{align}
\label{plus}
|(+,n)\rangle  =&  \cos\theta_n \, | e, n\rangle +
\sin\theta_n \, | g, n+1\rangle  \,, \\[2ex]
\label{minus}
|(-,n)\rangle  =&  -\sin\theta_n \, | e, n\rangle +
\cos\theta_n \, | g, n+1\rangle .
\end{align}
\end{subequations}
Here, $|i, n\rangle$ ($i\in \{e,g\}$) are  combined
atom-field states where the atom is in state $i$ 
with $n$ photons in the driving field mode.
The energies of these dressed states are given by
$E_{\pm,n} = (n+\nicefrac{1}{2})\omega_{\rm L} + 
\omega_{eg}/2 \pm \Omega_{\rm R}^{(n)}/2$,
where the splitting between the two dressed states $|(\pm, n)\rangle$
is known as the AC-Stark-shift.
The various spectral components then arise from 
transitions $|(\pm, n)\rangle \to |(\pm, n-1)\rangle$.
The transitions $+\to+$ and $-\to-$ yield the
central Mollow component and the coherent elastic
peak, while the transitions $+\to-$ and $-\to+$ yield
the sidebands shifted to higher and lower frequencies, respectively.

\section{\label{calculation}CALCULATION OF RELATIVISTIC AND RADIATIVE
CORRECTIONS TO THE MOLLOW SPECTRUM}

In the following, we discuss corrections whose understanding
is essential for the additional relativistic
and radiative energy shifts received by the dressed states.
First, we evaluate corrections which may be incorporated in a
redefinition of the detuning of the driving laser field to the
atomic transition frequency; 
in the second part, we complete
the analysis by considering corrections which effectively
modify the Rabi frequency.
Throughout the analysis, we focus on the 
hydrogen 1$S$-2$P$ transition as a promising
candidate for a possible experiment. 

In Sec.~\ref{results}, we have employed a purely nonrelativistic theory.
Both the resonance frequency as well as the transition dipole 
moments are evaluated first for the nonrelativistic (Schr\"{o}dinger) 
case. However, in order to resolve radiative effects, it is 
necessary to include the relativistic shifts of the transitions in a 
unified theory, and to analyze the fine-structure. 
The nonrelativistic expressions for the transition
dipole moments also change once we resolve the
fine-structure levels, because the angular momentum algebra is 
augmented by the spin.

%
%
\subsection{\label{sCorrDetuning}Corrections to the Detuning}

%
%
\subsubsection{\label{rel}Relativistic Corrections to the Resonance Frequency}

The well-known 
relativistic correction to the hydrogen energy levels is
given by
\begin{equation}
\label{Hrel}
H_{\rm rel} = - \frac{\bm{p}^4}{8 \, m^3} + 
\frac{\pi Z \alpha}{2 m^2} \, \delta(\bm{r}) + 
\frac{Z \alpha}{4 m^2 r^3} \, \bm{\sigma} \cdot \bm{L}\,.
\end{equation}
The effects are of the order of $(Z\alpha)^4\,m$, 
whereas the Schr\"{o}dinger energy is of the 
order of $(Z\alpha)^2\,m$. The 
full expression for the Dirac
energy of a hydrogenic level
with quantum numbers $n$, $j$, is~\cite[Eq.~(2-87)]{ItZu1980}
\begin{equation}
E_{nj} = m - \frac{(Z\alpha)^2\,m}{2 n^2} \\[2ex]
- \frac{(Z\alpha)^4 \, m}{n^3} \,  
\left[ \frac{1}{2 j + 1} - \frac{3}{8 \, n} \right] \,,
\end{equation}
where we neglect terms of order $(Z\alpha)^6$.
When evaluating the expectation values of $H_{\rm rel}$
on the dressed states (\ref{dressedstates}) in
first-order perturbation theory, the following
expression results,
\begin{equation}
\label{firstorderREL}
\delta \omega_{\pm,j}^{(\rm rel)} =
\mp \frac{\Delta}{\sqrt{\Omega^2 + \Delta^2}} \;
E_{\rm rel}^{(j)} \, .
\end{equation}
Here
\begin{subequations}
\begin{equation}
E_{\rm rel}^{(\nicefrac{1}{2})} = 
\left< 2 P_{\nicefrac{1}{2}} \left| H_{\rm rel} \right|
2 P_{\nicefrac{1}{2}} \right> -
\left< 1 S_{\nicefrac{1}{2}} \left| H_{\rm rel} \right|
1 S_{\nicefrac{1}{2}} \right>\,,
\end{equation}
and
\begin{equation}
E_{\rm rel}^{(\nicefrac{3}{2})} =
\left< 2 P_{\nicefrac{3}{2}} \left| H_{\rm rel} \right|
2 P_{\nicefrac{3}{2}} \right> -
\left< 1 S_{\nicefrac{1}{2}} \left| H_{\rm rel} \right|
1 S_{\nicefrac{1}{2}} \right>\,.
\end{equation}
\end{subequations}
The expression for $\delta \omega_{\pm,j}^{(\rm rel)}$ 
finds a natural interpretation as a first-order
(in $E_{\rm rel}^{(j)}$) correction to the quantity
\begin{eqnarray}
\lefteqn{\sqrt{\Omega^2 + \left(\Delta - E_{\rm rel}^{(j)}\right)^2}}
\nonumber\\
& & = \sqrt{\Omega^2 + \Delta^2}
- \frac{\Delta}{\sqrt{\Omega^2 + \Delta^2}} \;
E_{\rm rel}^{(j)} + \dots \,.
\end{eqnarray}
We can thus formally define a ``summed''  relativistic shift 
of the Mollow sidebands as
\begin{align}
\label{summedREL}
\Delta \overline{\omega}^{(\rm rel)}_{\pm,j} =
\pm \left( \sqrt{\Omega^2 + 
\left(\Delta - E_{\rm rel}^{(j)}\right)^2}
- \sqrt{\Omega^2 + \Delta^2}\right)\,.
\end{align}
We recall that the detuning has been defined as
$\Delta = \omega_{\rm L} - \omega_{eg}$ in Sec.~\ref{results}.
If the fine-structure is included, the resonance frequency
becomes $j$-dependent. The shift of the detuning 
as given in Eq.~(\ref{summedREL}) is thus equivalent to a
modification of the resonance frequency according to 
\begin{subequations}
\begin{eqnarray}
\omega_{eg} &\to& \omega^{(j)}_{eg} \equiv
\omega_{eg} + E_{\rm rel}^{(j)}\,,
\\[2ex]
\label{shiftdeltaREL}
\Delta &\to& \Delta - E_{\rm rel}^{(j)}\,.
\end{eqnarray}
\end{subequations}
Thus, the ``summed'' shift of the detuning
due to the relativistic 
correction $E_{\rm rel}^{(j)}$,
evaluated using the dressed-state basis,
is equivalent to the shift of the detuning
that would have been
obtained if we had evaluated the detuning, right from the 
start, with a resonance frequency corrected by the relativistic effects.
The ``summation'' implied by 
Eq.~(\ref{summedREL}) thus finds a natural interpretation.

Throughout the calculations, we will
refer to shifts of the Mollow sidebands $\omega_\pm$
due to first-order perturbations as
$\delta \omega_\pm$ [see e.g.~Eq.~(\ref{firstorderREL})], whereas summed 
expressions like~Eq.~(\ref{summedREL}) will be denoted
$\delta \overline{\omega}_\pm$.

%
%
\subsubsection{\label{Lamb}Bare Lamb Shift}

In addition to the relativistic shifts, the 
positions of the sidebands have to be modified further if
one desires a numerical accuracy as required 
to appropriately model current high-precision spectroscopy 
experiments. In~\cite{JeEvHaKe2003,JeKe2004},
the second-order radiative self-energy corrections
due to the interaction of the combined system of
atom and driving laser field with the surrounding non-laser-field
vacuum modes was analyzed. Taking into account both
interactions of the atom-field system with
resonant and off-resonant intermediate states,
in the limit $\Delta, \Omega \ll \omega_{eg}$ (i.e. 
under the replacements $\omega_{\rm L} \to \omega_{eg}$,
$\omega_{eg}-\omega_{\rm L} \pm \Omega_{\rm R} \to 0$, and
$\omega_{\rm L} + \omega_{eg} \pm \Omega_{\rm R} \to 2\omega_{eg}$) we 
obtain  corrections
to the energy of the dressed states which yield an 
additional shift of the position of the sidebands given by
\begin{equation}
\label{Deltaomegaplus}
\delta \omega_{\pm,j}^{(\rm Lamb)} = 
\mp \frac{\Delta}{\sqrt{\Omega^2 + \Delta^2}} \;
L_{\rm bare}^{(j)} \, .
\end{equation}
Here, the prefactor arises from the
mixing coefficients $\cos\theta$ and $\sin\theta$, and
$L_{\rm bare}^{(j)}$ is the usual Lamb shift of the atomic bare
state transition frequency which for the hydrogen 1$S$-2$P$ transition
is given by
\begin{eqnarray}
\label{Lbarej}
L_{\rm bare}^{(j)} &=& L_{2P_j} - L_{1S} \, ,
\end{eqnarray}
where $j=\nicefrac{1}{2},\nicefrac{3}{2}$ is the total 
angular momentum quantum number
of the excited state [for the definition of $L_{nl_j}$ 
see also Eq.~(\ref{defElamb}) below].
The Lamb shifts of the individual states are given by~\cite{JePa1996,PaJe2003}
\begin{eqnarray}
L_{1S} &=& 8172811(32)\: {\rm kHz} \, , \label{Lamb-1s}\\
L_{2P_{\nicefrac{1}{2}}} &=&  -12835.99(8)\: {\rm kHz}\, ,\\
L_{2P_{\nicefrac{3}{2}}} &=&  12517.46(8)\: {\rm kHz}\, .
\end{eqnarray}
The correction may be interpreted physically by defining
the dressed summed Lamb shift $\delta \overline{\omega}_{\pm,j}$ as
\begin{equation}
\label{summedLamb}
\delta \overline{\omega}^{(\rm Lamb)}_{\pm,j} = 
\pm \left( \sqrt{\Omega^2 + 
\left(\Delta - L_{\rm bare}^{(j)}\right)^2}
- \sqrt{\Omega^2 + \Delta^2}\right) \, ,
\end{equation}
where to first order in $L^{(j)}_{\rm bare}$, one recovers
Eq.~(\ref{Deltaomegaplus}).
Thus the correction $\delta \omega_{\pm,j}^{(\rm Lamb)}$ 
effectively is a shift 
\begin{equation}
\Delta \to \Delta - L_{\rm bare}^{(j)}
\end{equation}
of the detuning [in analogy to (\ref{shiftdeltaREL})].

With typical parameters
(see Sec. \ref{num-analysis}), the summed expression 
$\delta \overline{\omega}_\pm^{(j)} $ yields results which
significantly differ from the first-order expression 
$\delta \omega_{\pm,j}^{(\rm Lamb)}$. The reason is 
that the bare Lamb shift is not small as compared to the 
detuning $\Delta$, so that the higher-order
terms of the series expansion are relevant.  
Nevertheless we use the summed formula Eq.~(\ref{summedLamb}) 
instead of Eq.~(\ref{Deltaomegaplus}), as it is the 
expected result in the sense that the Lamb shift is 
naturally interpreted as a 
modification of the transition frequency and a 
corresponding alteration of the detuning.

%
%
\subsubsection{\label{unified}Unified Expressions for the 
Relativistic and Radiative Shifts}

Both the summed relativistic shift Eq.~(\ref{summedREL})
as well as the summed Lamb shift Eq.~(\ref{summedLamb}) 
are effectively summarizing the corrections 
received by the detuning due to various 
shifts that go beyond the nonrelativistic treatment
of the hydrogen (and Mollow) spectrum discussed in Sec.~\ref{results}.
These effects would also be observable in low-intensity
scattering of (laser) light off atoms, and are
automatically included in the observable resonance frequency
of the transition whose high-intensity behavior
we are studying. The corrections can therefore
be included into the formalism if we replace the detuning
$\Delta$ by the detuning $\Delta_{\rm exp}$ to the experimental 
transition frequency given by
\begin{equation}
\label{redefDelta}
\Delta_{\rm exp} = \omega_{\rm L} - \omega_{\rm exp}\, ,
\end{equation}
where $\omega_{\rm exp}$ is the experimentally 
observable transition frequency as it would be 
obtained from low-intensity scattering~\cite{KrHe1925}.
In Sec.~\ref{results}, we have started from 
a nonrelativistic theory, and therefore the detuning
$\Delta = \omega_{\rm L} - \omega_{eg}$ was calculated
with regard to the inaccurate resonance frequency $\omega_{eg}$
as it follows from the Schr\"odinger theory that 
fails, as is well known, to describe even the relativistic
effects that lead to the fine-structure (let alone the Lamb shift).
Thus, in practice, the bare Lamb shift modification to the detuning 
may be accounted for by replacing the 
resonance frequency $\omega_{eg}$ as it would 
be obtained from a nonrelativistic theory,
by an experimental value for the
atomic transition frequency as found in low intensity
scattering experiments~\cite{KrHe1925}.

The frequency $\omega_{\rm exp}$ may not be known 
well enough for any given transition to lead to 
a meaningful comparison between theory and a 
conceivable high-accuracy measurement of the 
Mollow spectrum. This is because we are sensitive, in 
the measurement of the Mollow spectrum, to tiny differences
between the laser frequency and the actual resonance 
frequency. It may therefore be useful to recall that 
for the Lamb shift $L_{nl_j}$ of a hydrogenic energy 
level (spectroscopic notation $nl_j$),
one may use the implicit definition 
(see e.g.~\cite{SaYe1990,JePa1996}),
\begin{eqnarray}
\label{defElamb}
\lefteqn{E(nl_j) = m_r \left[ f(n,j)-1 \right]} \nonumber\\
& & - \frac{m_r^2}{2 (m + m_N)}
\left[ f(n,j) - 1 \right]^2 + L_{nl_j} + E_{\rm hfs}\,,
\end{eqnarray}
where $E$ is the energy level of the two-body-system
and $f(n,j)$ is the dimensionless Dirac energy, $m$
is the electron mass, $m_r$ is the reduced mass of the system
and $m_N$ is the nuclear mass. In very accurate
experiments, one also has to include the hyperfine frequency 
shift $E_{\rm hfs}$ which depends on the quantum number
$m_F$ that includes the nuclear spin.
Note, however, that the hyperfine structure does not contribute to the
Lamb shift according to the definition Eq.~(\ref{defElamb}).

The expression Eq.~(\ref{defElamb}) can be used to make a
theoretical prediction $\omega_{\rm th}$ for the transition frequency by
forming the difference of this expression for the two states
involved in the atomic transition whose high-intensity
behavior we are studying. The detuning can then 
alternatively be evaluated as 
$\Delta_{\rm th} = \omega_{\rm L} - \omega_{\rm th}$.
Assuming $\omega_{\rm th} = \omega_{\rm exp}$, one then has
$\Delta_{\rm exp} = \Delta_{\rm th}$.
For some recent data on Lamb shifts, we
refer to~\cite{PaJe2003}.

%
%
\subsubsection{\label{bloch}Bloch--Siegert shifts}

The Mollow spectrum also
receives corrections due to so-called counter-rotating
interactions of the driving laser field with the atom~\cite{BlSi1940}.
These correspond to an excitation of the atom simultaneously
with a creation of a laser photon or the vice-versa
process. The first-order perturbation vanishes, and the
second-order expression is given by~\cite{JeKe2004} 
\begin{eqnarray}
\label{corrBS}
\Delta E^{(\rm BS)}_{\pm} &=& 
\pm \frac{\Omega^2}{\omega_{\rm L}} \, 
\frac{8 \, \cos(2\theta) - (\Omega_{\rm R}/\omega_{\rm L}) \,
\left[3 + \cos(4 \theta) \right]}
  {64 - 16 \, (\Omega_{\rm R}/\omega_{\rm L})^2}
\nonumber\\[2ex]
&=& \pm \frac18\,
\frac{\Omega^2}{ \sqrt{\Omega^2 + \Delta^2}}\,
\frac{2 \Delta^2 + \Omega^2 + 4 \, \Delta \, \omega_{\rm L}}
  {\Delta^2 + \Omega^2 - 4 \, \omega^2_{\rm L}}  \, .
\end{eqnarray}
The correction $\delta \omega^{(\rm BS)}_\pm$ of the Mollow sidebands 
due to the Bloch-Siegert shift is thus given by
\begin{eqnarray}
\delta \omega^{(\rm BS)}_\pm &=& 
\Delta E^{(\rm BS)}_{\pm} -  \Delta E^{(\rm BS)}_{\mp} \\
&=& \mp \frac{\Delta}{ \sqrt{\Omega^2 + \Delta^2}}\,
[{\mathcal B}\, \Omega^2] \,. \label{shift-b}
\end{eqnarray}
Here, the parameter $[{\mathcal B}\, \Omega^2]$ depends
on the laser field intensity, which reflects the
fact that the Bloch-Siegert shifts are stimulated processes. 
Assuming $\Omega, \Delta \ll \omega_{\rm L}$ as a
typical range of parameters, one 
has $\omega_{\rm L} \approx \omega_{\rm R}$ and thus  
\begin{equation}
{\mathcal B} = \frac{1}{4\, \omega_{\rm R}} + 
{\mathcal O}(\Delta/\omega_{\rm L}^2, \Omega/\omega_{\rm L}^2) \,.  
\label{par-b}
\end{equation}
Here one should note that the energy shift Eq.~(\ref{shift-b}) with
${\mathcal B}$ as in Eq.~(\ref{par-b}) vanishes for $\Delta = 0$.
This is consistent with the analysis in \cite{BrKe2000,JeKe2004}, where it
was found that the Bloch-Siegert shift is suppressed by an additional
power of $\Omega / \omega_{\rm L}$ for vanishing detuning.
We also define a summed
Bloch-Siegert shift in analogy with (\ref{summedREL}) 
and (\ref{summedLamb}) as
\begin{equation}
\label{summedBS}
\delta \overline{\omega}^{(\rm BS)}_{\pm} =
\pm \left( \sqrt{\Omega^2 + (\Delta - {\mathcal B}\, \Omega^2)^2}
- \sqrt{\Omega^2 + \Delta^2}\right)\,.
\end{equation}
Effectively, the Bloch--Siegert shift may be accounted for by the replacement
$\Delta \to \Delta - {\mathcal B}\, \Omega^2$. This correction 
to the detuning is proportional to $\Omega^2$, i.e.~proportional
to the laser intensity.

%
%
\subsubsection{\label{sec-offres}Off--Resonant Radiative Corrections
Stimulated by the Atom--Laser Interaction}

For these corrections, we restrict the atom-field interaction 
to the laser mode, but take into account the off-resonant (OR)
atomic levels $|j\rangle$ (i.e. $|j\rangle \neq |e\rangle, |g\rangle$).
The leading effect is the second-order perturbation
\begin{eqnarray}
\label{stimradcor}
& & \Delta E^{(\rm OR)}_{\pm,n} = 
\\[2ex]
& & \left< (\pm, n) \left| {\mathcal H}_{\rm L} \,
\frac{1}{E_{\pm,n} - 
\left( {\mathcal H}_{\rm M} + {\mathcal H}_{\rm F} \right) } \,
{\mathcal H}_{\rm L} \right| (\pm,n)  \right>\,.
\nonumber
\end{eqnarray}
Here, we have defined the Schr\"odinger-picture Hamiltonian 
${\cal H}_{\rm L}$ describing 
the interaction of the atom with the driving
laser field, the free energy of the nonresonant
atomic states ${\cal H}_{\rm M}$, and the free energy of the 
electromagnetic fields ${\cal H}_{\rm F}$
as
\begin{subequations}
\begin{eqnarray}
{\cal H}_{\rm L} &=& -q \, {\bm x} \cdot {\bm E}_{\rm L} \, ,\\
{\cal H}_{\rm M} &=& 
\sum_{j \neq g,e} \omega_j \, | j \rangle \, \langle j | \, ,\\
{\cal H}_{\rm F} &=& \sum_{{\bf k}\lambda} \omega_{{\bf k}} \,
a^+_{{\bf k}\lambda} \, a_{{\bf k}\lambda} \, , \label{ham-l}
\end{eqnarray}
\end{subequations}
respectively, where $a_{{\bf k}\lambda}$ and 
$a^+_{{\bf k}\lambda}$ are annihilation and
creation operators for photons with wave 
vector ${\bf k}$, frequency $\omega_{{\bf k}}$
and polarization $\lambda$, and $\omega_j$ 
($j \neq g,e$) are the energies of the nonresonant 
intermediate states. The sum in Eq.~(\ref{ham-l})
extends over all possible vacuum field modes, and 
${\bm E}_{\rm L}$ is the field operator for the 
laser mode,
\begin{equation}
{\bm E}_{\rm L} = \sqrt{\frac{\omega_{\rm L}}{2 \, V}}\,
\bm{\epsilon}_{\rm L}\,
\left[ a_{\rm L} + a^+_{\rm L} \right].
 \end{equation}
Here, $\bm{\epsilon}_{\rm L}$ is the polarization vector
for the laser mode.
As the laser mode is highly populated with 
an occupation number $n \gg 1$, both the field
annihilation and creation operators in ${\mathcal H}_{\rm L}$ contribute
[see also Eq.~(\ref{matching})].
The resulting expression for the energy shift of the dressed state
$|(+,n)\rangle$ is given by
\begin{eqnarray}
\label{stimradplus}
&&\Delta E^{(\rm OR)}_{+,n} = \sum_{j \neq e,g} \bigg\{ 
|g_{ej}|^2 \, \cos^2\theta \nonumber\\[2ex]
&& \times
\left( \frac{n+1}{-\omega_j + E_1} + \frac{n}{-\omega_j + E_2}\right) 
+ |g_{gj}|^2 \, \sin^2\theta \,
\nonumber\\[2ex]
& & \times
\left( \frac{n+1}{-\omega_j + E_3} + 
\frac{n}{-\omega_j + E_4}\right) \bigg\}\,.
\end{eqnarray}
with  
\begin{subequations}
\label{energies-d}
\begin{eqnarray}
E_1 &=& \omega_g - \frac12\,\Delta + \frac12\, \Omega_{\rm R} 
\nonumber\\
&\approx& \omega_g \, ,\\
E_2 &=& \omega_g+ 2\,\omega_{\rm R} + 
  \frac32\,\Delta + \frac12\, \Omega_{\rm R}  
\nonumber\\
&\approx& 
\omega_g+ 2\,\omega_{\rm R} \, ,\\
E_3 &=& \omega_g - \omega_{\rm R} - 
  \frac32\,\Delta + \frac12\, \Omega_{\rm R}  
\nonumber\\
&\approx& 
\omega_g - \omega_{\rm R} \, ,\\
E_4 &=& \omega_g+ \omega_{\rm R} + 
  \frac12\,\Delta + \frac12\, \Omega_{\rm R}  
\nonumber\\
&\approx& 
\omega_g+ \omega_{\rm R} \, .
\end{eqnarray}
\end{subequations}
For the dressed state $|(-,n)\rangle$, we have
\begin{eqnarray}
\label{stimradminus}
\lefteqn{\Delta E^{(\rm OR)}_{-,n} = \sum_{j \neq e,g} \bigg\{
|g_{ej}|^2 \, \sin^2\theta}
\nonumber\\[2ex]
& & \times \left( 
\frac{n+1}{-\omega_j + E_1} + 
\frac{n}{-\omega_j + E_2}\right) +
|g_{gj}|^2 \, \cos^2\theta 
\nonumber\\[2ex]
& & \times \left( 
\frac{n+1}{-\omega_j + E_3}  + 
\frac{n}{-\omega_j + E_4} \right ) 
\bigg\} \,. 
\end{eqnarray}
In calculating these expressions, we may carry 
out the semiclassical approximation
$n + 1 \approx n$ and replace the above Rabi frequency by its semiclassical
counterpart. The coupling $g_{ij}$ is defined by
\begin{equation}
g_{ij} = 
- q \, \langle i | \bm{\epsilon}_{\rm L} \cdot \bm{x} | j \rangle \,
\sqrt{\frac{\omega_{\rm L}}{2 \, V}}\,,
\end{equation}
and is of the same order of of magnitude as $g_{\rm L}$
We therefore obtain as the second-order shift due to  the 
off-resonant energy levels,
\begin{eqnarray}
\label{stimradplusminus}
\lefteqn{\delta \omega^{\rm (OR)}_\pm =
\Delta E^{(\rm OR)}_{\pm,n} - \Delta E^{(\rm OR)}_{\mp,n}}
\nonumber\\[2ex]
&=& \pm {\mathcal D}\, \cos(2\theta) \, \Omega^2 \, 
= \mp \frac{\Delta}{\sqrt{\Omega^2 + \Delta^2}} \,
[{\mathcal D}\, \Omega^2] \,, 
\end{eqnarray}
where $\Delta$ is the detuning and 
\begin{eqnarray}
{\mathcal D} &=& \frac{1}{4\,g^2_{\rm L}} \, 
\sum_{j \neq e,g} \left\{
|g_{ej}|^2 \, 
\left ( \frac{1}{-\omega_j + E_1} + 
\frac{1}{-\omega_j + E_2} \right ) \right. \nonumber \\
&& \qquad  \left . - |g_{gj}|^2  \,
\left ( \frac{1}{-\omega_j + E_3} + 
\frac{1}{-\omega_j + E_4} \right )  \right \} 
\end{eqnarray}
depends again on the laser intensity. 
The energies $E_i$ ($i=1,\dots,4$)
are defined in Eq.~(\ref{energies-d}).
Therefore this  additional shift is a 
stimulated radiative correction in the same sense as the corrections
discussed in the previous section~\cite{CTDRGr1992}. 

To further evaluate the parameter ${\mathcal D}$, it is important
to note that the virtual states are coupled to the initial states
by the driving laser field rather than by the vacuum as for example in 
Lamb shift calculations. Thus the polarization of the coupling field
mode is fixed. For the off-resonant corrections,
it is sufficient to evaluate the relevant matrix elements
in the nonrelativistic approximation.
So, if we assume the atoms to be in the 
1$S$ ground state initially ($m=0$, of course), then we have a situation
in which for a given polarization of the
laser field not all magnetic sublevels of the 
2$P$ states are coupled. In the following, we assume the driving
laser field to be $z$-polarized [${\bm  \epsilon}_{\rm L} = (0,0,1)$], 
so that only the $m=0$ sublevels
of the 1$S$ and the 2$P$ ground and excited state are occupied.
Then the parameter ${\mathcal D}$ may be rewritten as 
[see Eq.~(\ref{energies-d})]
\begin{align}
{\mathcal D} = \frac{M_g(E_3) + M_g(E_4) - M_e(E_1) - M_e(E_2)}
{4 \left | \left <g|z|e\right > \right |^2}
\end{align}
in terms of the two matrix elements
\begin{eqnarray}
M_g(\zeta) &=&  \left< g \left| \, z  
\, G^{''}(\zeta) \, z\, \right| g \right>  \label{m1}\, ,\\
M_e(\zeta) &=&  \left< e \left| \, z\,  G^{''}(\zeta) \, z \,
 \right| e \right> \, , \label{m2}
\end{eqnarray}
where the propagator is given by 
\begin{equation}
G(\zeta) = \frac{1}{H - \zeta}\,,
\end{equation}
and where the double prime means that {\em both} resonant states 
$|g\rangle = |1S, m\!\!=\!\!0\rangle$ and 
$|e\rangle = |2P, m\!\!=\!\!0\rangle$ 
are {\em excluded} from the sum over intermediate states
in the Green function.
The evaluation of $M_e(\zeta)$ requires special care, as there are 
both $S$ (angular quantum number $l=0$) and $D$ ($l=2$)
states as intermediate states. 
Due to the fixed polarization of the coupling field, the angular parts 
of these two contributions have different proportionality 
factors relative to the 
angular parts of the ``standard'' matrix element 
\begin{equation}
\sum_{i=1}^{3} \left< e \left| x^i \,
G^{''}(\zeta) \, x^i \right| e \right>
\end{equation}
and thus have to be calculated separately. The above standard matrix element
may however be recovered from the  matrix elements for definite initial state
and coupling field polarization by averaging appropriately.
This is discussed in Appendix~\ref{app-derivation}.

In the propagator $G(\zeta)$, the energy is parameterized by the
(in general complex) parameters
\begin{subequations}
\label{tparam}
\begin{eqnarray}
\zeta &\equiv& \zeta(t) = -\frac{(Z\alpha)^2 \, m}{2 \, n^2 \, t^2}\,,\\[2ex]
t &\equiv& t(\zeta) = \frac{Z\alpha}{n} \,
\sqrt{-\frac{m}{2\zeta}}\,.
\end{eqnarray}
\end{subequations}
Usually, one has $\zeta = E - \omega$ where $E$ is the bound state energy,
and we may write
\begin{equation}
\zeta(t) =  E - \omega(t) =
- \frac{(Z\alpha)^2 \, m}{2 \, n^2} - \omega(t) \,.
\end{equation}
The parameters $\omega$ and $t$ are related by the equations
\begin{subequations}
\begin{eqnarray}
\label{ttoomega}
\omega &=& \frac{(Z\alpha)^2 m}{2 n^2}\,\frac{1-t^2}{t^2}\,,
\\[2ex]
t_{nl}(\omega) &=& 
\left(1 + \frac{2 n^2 \omega}{m (Z\alpha)^2}\right)^{-1/2}\,,
\end{eqnarray}
\end{subequations}
where $n,l$ are the principal and the angular momentum quantum number
of the quantum state for which the relevant matrix elements 
are to be evaluated. In the following,
we will also use common spectroscopic notation
for the level characterized by $n$ and $l$, 
i.e. for example~$t_{2P}(\omega) \equiv t_{21}(\omega)$.
For the energies $E_i$ ($i\in \{1,\dots,4\}$) and $n=1,2$ we thus obtain
\begin{equation}
t_{nl}(E_i) = \sqrt{\frac{E_n}{E_i}}\, .
\end{equation}
The above matrix elements Eqs.~(\ref{m1}, \ref{m2}) without the double 
primes, i.e. including resonant intermediate states, may then be expressed
in terms of the ``standard'' hypergeometric 
function~\cite{Pa1993,JePa1996,JeSoMo1997} 
\begin{equation}
\Phi(n, t) =  {}_2F_1\left(1, - n t, 1 - n t, 
\left( \frac{1-t}{1+t} \right)^2 \right)\,.
\end{equation}
as (see Appendix \ref{app-derivation})
\begin{subequations}
\begin{align}
\overline{M}_g(\zeta) =& \left< g \left| \, z\, 
G(\zeta) \, z\, \right| g \right> =
m\:a^4_{\rm B}\,  \left[ 
\frac{2\,t^2\,{\cal X}_g(t)}
  {3\,(t - 1)^5 \, (t + 1)^4} \right .\nonumber \\
&\left . - \frac{256\,t^9}{3 \, (t - 1)^5 \, (t + 1)^5} \, \Phi(1, t) \right]\,, \label{matrix1s} \\
\overline{M}_e(\zeta) =& \left< e\left | \, z \,
G(\zeta) \,z \,  \right| e \right> =
m\:a^4_{\rm B}\,\left[
\frac{16\,t^2\,{\cal X}_e(t)}
  {15\,(t - 1)^7 \, (t + 1)^5} \right. \nonumber \\
&\left. -\frac{2^{14}\,t^{11}\,(23 t^2 - 7)}{15\,(t-1)^7\,(t+1)^7}\,
\Phi(2, t) \right]\,, \label{matrix2p}
\end{align}
\end{subequations}
where 
\begin{align}
{\cal X}_g(t)  =&\:  38  t^7 + 26 t^6 + 19 t^5 -
19 t^4 - 12 t^3 \nonumber \\
& + 12 t^2 + 3 t - 3 \,,\\
{\cal X}_e(t)  =&\:  6739 t^{10} - 1702 t^9 - 231 t^8 - 1420 t^7
\nonumber \\
& - 262 t^6 + 1944 t^5 - 402 t^4 - 1140 t^3 
\nonumber \\
& + 435 t^2 + 270 t - 135\,.
\end{align}
Here, the Bohr radius scaled by the nuclear charge number $Z$, 
in our units, is
\begin{equation}
\label{aBohr}
a_{\rm B} = \frac{1}{Z\alpha m}\,,
\end{equation}
where $\alpha$ is the fine-structure constant and $m$ is the 
electron mass.

The corresponding matrix elements without resonant intermediate states
may then be obtained by subtracting the respective contributions
of the resonant intermediate states
\begin{eqnarray}
M_g(\zeta) &=& \left< g \left| \, z\, 
G^{''}(\zeta) \, z\,  \right| g \right> \nonumber \\
&=& \overline{M}_g(\zeta) - 
\frac{\left | \left< g \left| \, z\,  \right | e \right > \right |^2}
  {E_{2P} - \zeta} \, ,\\
M_e(\zeta) &=& \left< e \left| \, z\, 
G^{''}(\zeta) \, z \,  \right| e \right> \nonumber\\
&=& \overline{M}_e(\zeta) - \frac{\left | \left< g \left|\, z\, 
 \right | e \right > \right |^2}{E_{1S} - \zeta } \, .
\label{cancellation}
\end{eqnarray}
We assume here that the Rabi frequency 
is not excessively large, which implies that it is small
as compared to the optical transition frequency 
(i.e.,~$\Omega \ll \omega_{\rm R}$).
For a meaningful measurement of the Mollow spectrum, 
it is necessary, furthermore, to tune the laser 
close to the atomic resonance
(which implies $\Delta \ll \omega_{\rm R}$).
In this case we 
may carry out the following approximations
[c.f. Eq.~(\ref{energies-d})]
\begin{subequations}
\label{t-d}
\begin{align}
&E_1 \to \omega_g  & \Rightarrow \qquad & t_{2P}(E_1) \to 1/2 \, ,\\
&E_2 \to \omega_g + 2 \omega_{\rm R}  & 
\Rightarrow \qquad & t_{2P}(E_2) \to \sqrt{-0.5} \, , \\
&E_3 \to \omega_g - \omega_{\rm R}  & 
\Rightarrow \qquad & t_{1S}(E_3) \to 2/\sqrt{7} \, , \\
&E_4 \to \omega_g + \omega_{\rm R}  & 
\Rightarrow \qquad & t_{1S}(E_4) \to 2 \, .
\end{align}
\end{subequations}
With these parameters, $\mathcal D$ evaluates to
\begin{eqnarray}
{\mathcal D} &=& \frac{1}{(Z\alpha)^2 m} \left 
[ 6.2148(8) - 0.23532(2)\, {\rm i} \right ] \nonumber \\
&=& \frac{1}{\omega_{\rm R}} \,
\left[ 2.3305(3) - 0.088245(6)\, {\rm i} \right] \, .
\end{eqnarray}
The uncertainties are mainly due to the approximations
carried out in Eq.~(\ref{t-d}) with respect
to the energies $E_i$ ($i = 1,\dots,4$)
originally defined in Eq.~(\ref{energies-d}). 
This is possible because the off-resonant 
stimulated radiative correction amounts to a modification
of the detuning which is 
of order $\Omega^2/\omega_{\rm R} \ll \Omega$ 
[see also Eq.~(\ref{summedOR}) below].
Therefore we may carry out the approximation
Eq.~(\ref{t-d}), i.e. neglect
the further corrections of order 
$\Omega^3/\omega_{\rm R}^2 \ll \Omega^2/\omega_{\rm R}$ 
and $\Omega^2 \Delta/\omega_{\rm R}^2 \ll \Omega^2/\omega_{\rm R}$,
which are beyond the scope of the current analysis.
The real part ${\mathcal D}_{\rm R} \equiv {\rm Re}\: ({\mathcal D})$
gives rise to a shift of the position of the Mollow sideband, 
while the imaginary part ${\mathcal D}_{\rm I} \equiv {\rm Im}\: ({\mathcal D})$
describes the ionization into the continuum. 
This means that the imaginary part of the energy shifts
$\Delta E^{(\rm OR)}_{\pm,n}$
received by the two dressed states $|(\pm, n)\rangle$
must be negative, which is equivalent to a 
negative sign for the imaginary part ${\mathcal D}_{\rm I}$.

In the numerical analysis in Sec.~\ref{num-analysis} it is shown that for 
typical parameters the system is sufficiently far
from the ionization threshold~\cite{ShBe1990} so that
the ionization does not restrict the applicability of our scheme.
The real part yields a correction
to the detuning given by 
$\Delta \to \Delta - {\mathcal D}_{\rm R}\Omega^2$,
according to the summation 
[cf. Eq.~(\ref{stimradplusminus})]
\begin{equation}
\label{summedOR} 
\delta \overline{\omega}^{\rm (OR)}_\pm =
\pm \left( \sqrt{\Omega^2 + (\Delta - {\mathcal D}_{\rm R}\Omega^2)^2}
- \sqrt{\Omega^2 + \Delta^2}\right)\,.
\end{equation}
with 
\begin{equation}
\delta \overline{\omega}^{\rm (OR)}_\pm \approx
\delta \omega^{\rm (OR)}_\pm = 
\mp \frac{\Delta}{\sqrt{\Omega^2 + \Delta^2}} \,
[{\mathcal D}_{\rm R}\, \Omega^2]\,.
\end{equation}
As pointed out below in section~\ref{num-analysis}
(see also Tab.~\ref{tabular}),
the magnitude of the off-resonant correction is 
small as compared to the detuning for typical parameters so that 
there is no numerically significant difference between
the first-order correction Eq.~(\ref{stimradplusminus}) 
and the summed form given in Eq.~(\ref{summedOR}).

The first-order imaginary contribution to the 
Mollow sidebands is 
\begin{equation}
\label{omegaIm}
\delta \omega^{\rm (Im)}_\pm =
\mp \frac{\Delta}{\sqrt{\Omega^2 + \Delta^2}} \,
[{\rm i}\,{\mathcal D}_{\rm I}\, \Omega^2]\,.
\end{equation}
This effect broadens the sidebands slightly, but 
its contribution is so small for typical parameters
(see Sec.~\ref{num-analysis} below), that it may
be ignored on the current level of accuracy of
the theoretical predictions.
It is interesting to note that the stimulated 
off-resonant correction is small as compared to the relatively 
large effect mediated by the bare Lamb shift of the transition,
which is discussed in Sec.~\ref{Lamb}.

%
%
\subsection{\label{sCorrRabi}Corrections to the Rabi Frequency}

In this section, we discuss 
corrections to the incoherent fluorescence
spectrum due to modifications of the Rabi frequency.
In particular, we consider corrections
to the transition dipole moment and to the
spontaneous transition rate of the atomic transition,
due to coupling of the driving 
laser field to resonant and nonresonant atomic transitions,
and corrections to the secular approximation.

\subsubsection{\label{rel-dipole}Relativistic Corrections to the 
Transition Dipole Moment}

In this section we discuss relativistic corrections
to the fluorescence spectrum up to relative order $(Z\alpha)^2$. 
The corrections amount to a modification of the 
atomic transition frequency and of the transition
dipole moments. The relativistic expressions
for the state energies and the transition dipole moments 
depend on the total angular
quantum numbers $j$ of the involved states,
which is the vectorial sum of the electron orbital 
angular momentum $l$ and its spin $s$.
Therefore we specify the total angular momentum quantum number and 
thus the spin state of the atomic system in order to  
fix a specific experimental setup (see also Appendix~\ref{app-dipole}).
We further assume the atom to be driven by a pure dipole laser field
linearly polarized in $z$-direction, such that the laser field only 
couples states with equal magnetic quantum number. The situation 
of a pure dipole field has recently
been studied in a related context in~\cite{Ya2003}.
In the numerical analysis in Sec.~\ref{num-analysis}, we consider a 
standing wave laser field configuration where the atom is at a 
point of maximum electric field intensity of the standing wave. As then the 
magnetic field component of the driving laser may be neglected, it 
is not considered in the following analysis (corrections due to 
the variation of the electric field about its maximum are treated in 
Sec.~\ref{field-config}).

The relativistic corrections to the energies of the 
atomic states and thus to the atomic transition frequencies
effectively modify the detuning $\Delta$ and may be
accounted for by choosing an experimental value for the atomic
transition frequency as found in low-intensity scattering
experiments (see Sec.~\ref{unified}).
The corrections to the transition dipole moments
may be evaluated with the help of
the relativistic wavefunctions of the hydrogen atom
as given in~\cite{SwDr1991a,SwDr1991b}. 
We denote the absolute relativistic correction to the 
nonrelativistic matrix element
\begin{equation}
\langle 1S_{\nicefrac{1}{2}}, m\!=\!\pm \nicefrac{1}{2} | \: z \: 
| 2P_{j}, m\!= \!\pm \nicefrac{1}{2} \rangle_{\textrm {NR}}
\end{equation}
by $\delta d_{j}^{(\rm R)}$,
where $j=\nicefrac{1}{2}, \nicefrac{3}{2}$ is the total angular momentum and  
$m =\pm  \nicefrac{1}{2}$ is the magnetic quantum number.
In the following, we will omit the ``$m=$'' from the second
parameter of the atomic state vectors.
Then the relativistic matrix element
(with subindex ``R'') gives rise to a relative
$(Z\alpha)^2$-correction $\delta d_{j}^{(\rm R)}/d_{j}$
with respect to the nonrelativistic (NR) expression
which is given by 
\begin{eqnarray}
\label{defdeltadj}
\frac{\langle 1S_{\nicefrac{1}{2}},\pm \nicefrac{1}{2} | \: z \: | 
2P_{j}, \pm \nicefrac{1}{2} \rangle_{\textrm R}}
{\langle 1S_{\nicefrac{1}{2}},\pm  \nicefrac{1}{2} | \: z \: | 
2P_{j},\pm  \nicefrac{1}{2} \rangle_{\textrm {NR}}} =  
1+\frac{\delta d_{j}^{(\rm R)}}{d_{j}} \,,
\end{eqnarray}
where we ignore higher-order relativistic terms of order $(Z\alpha)^m$
with $m \geq 4$. 
The corresponding matrix elements where the 
``initial'' and the ``final'' state
have different magnetic quantum numbers  vanish identically
as the driving laser field is assumed to be polarized in the $z$-direction.
Evaluating the relative corrections, one obtains
\begin{subequations}
\label{deltarREL}
\begin{align}
\frac{\delta d_{\nicefrac{1}{2}}^{(\rm R)}}{d_{\nicefrac{1}{2}}} =& 
-(Z\alpha)^2 \left(\frac{13}{32}+
\frac{3}{2}\ln 2 - \ln 3 \right) \, ,\\
\frac{\delta d_{\nicefrac{3}{2}}^{(\rm R)}}{d_{\nicefrac{3}{2}}} =& 
-(Z\alpha)^2\left(\frac{31}{96}+
\frac{5}{4}\ln 2 - \frac{3}{4} \ln 3 \right) \, .
\end{align}
\end{subequations}
The Rabi frequency and the transition dipole moment depend
linearly on each other. Therefore the relative correction to the 
Rabi frequency
is identical to the relative correction to the transition dipole moment:
\begin{equation}
\frac{\delta \Omega^{(\rm R)}_j}{\Omega} = 
\frac{\delta d_j^{(\rm R)}}{d_j} \, . \label{relom}
\end{equation}
Here, $\delta \Omega^{(\rm R)}_j$ is the absolute correction to 
the Rabi frequency due to relativistic modifications to the transition
dipole moment. The spin-dependent shift 
$\delta \overline{\omega}^{(\rm R)}_{\pm,j}$
of the position of the Mollow sidebands due to 
the relativistic corrections
of the matrix element is thus given by:
\begin{equation}
\label{summedR}
\delta \overline{\omega}^{(\rm R)}_{\pm,j} =
\pm \left ( \sqrt{\left (\Omega+ 
\delta \Omega^{(\rm R)}_j\right )^2 + \Delta^2} 
- \sqrt{\Omega^2 + \Delta^2} \right ) \,,
\end{equation}
where
\begin{equation}
\delta \overline{\omega}^{(\rm R)}_{\pm,j} 
\approx \delta \omega^{(\rm R)}_{\pm,j} 
= \mp \frac{\Omega^2}{\sqrt{\Omega^2 + \Delta^2}} \, 
{\mathcal E_{j}} 
\end{equation}
and
\begin{equation}
{\mathcal E_{j}} = 
- \frac{\delta d_{j}^{(\rm R)}}{d_j} + 
{\mathcal O}\left ( \frac{\delta d_{j}}{d_j}\right )^2 \, .
\end{equation}
With these definitions, the summed relativistic correction to the dipole
moment effectively corresponds to a replacement
$\Omega \to \Omega (1-{\mathcal E_{j}})$.

\subsubsection{\label{field-config}Field--Configuration Dependent 
Correction to the Rabi Frequency}

It is well-known that the magnetic component of plane-wave electromagnetic
wave influences the transition current at relative order 
$(Z\alpha)^2$ (see e.g.~\cite{Je1996,JePa1996,Pa2004}). 
The radiation pressure due to the magnetic field 
could move the atom. Therefore, we propose a standing-wave
field configuration, where the atom is placed at an anti-node
of the standing-wave electric field. In this setup, the influence
of the magnetic field can be neglected to a very good approximation.

The analysis of the previous Sec.~\ref{rel-dipole}
is valid up to the order discussed (relative order $(Z\alpha)^2$) 
only for a pure dipole field which additionally has to be constant 
in any direction  perpendicular to the polarization.
However, for a standing wave configuration, the $z$-polarized 
electric field of the laser is not constant in the propagation 
($x$-)direction. This leads to a further correction, which 
gives rise to a field-configuration dependent shift of the Rabi 
frequency. In the following, this shift of
relative order $(Z\alpha)^2$ is analyzed 
for the setup described above where the atom is at the maximum of 
the standing-wave electric field.

We start from the long-wavelength quantum electrodynamic
(LWQED) interaction Hamiltonian~\cite{Pa2004}.
The only relevant terms (in the context of our 
analysis) of the interaction part of this Hamiltonian are 
\begin{eqnarray}
\label{lwqed}
H_I^{\rm{LW}} &=& 
-q\,\bm{x}\cdot\bm{E}
-\frac{q}{2}\,x^i\,x^j\,E^i_{,j} \nonumber \\
&&-\frac{q}{6}\,x^i\,x^j\,x^k\,E^i_{,jk} \,.
\end{eqnarray}
Here, the $x^i$ denotes the $i$th component of 
the position operator vector $\bm x$, and $E^i_{,j}$
is the partial derivative with respect to $x^j$ of the
$i$th component of the electric field vector.
The electric field of the standing wave is given by
\begin{equation}
\bm{E}(t, x) = \hat{e}_z \, {\mathcal E}_{\rm{SW}} 
  \, \cos(\omega t) \, \cos(k x)\,.
\end{equation}
The term containing the first derivative of the electric field
in Eq.~(\ref{lwqed}) vanishes, and the last term gives
\begin{eqnarray}
-\frac{q}{6}\,x^i\,x^j\,x^k\,E^i_{,jk} 
&=&-\frac{q}{6}\,z\, x^2\,E^z_{,xx} \nonumber \\
&=&\frac{q}{6} \, z \, k^2 \, x^2 \, 
 {\mathcal E}_{\rm{SW}} \, \cos(\omega t) \,. \label{last-term}
\end{eqnarray}
This result has to be distinguished from a simple expansion of 
the electric field around the maximum at $x=0$,
which yields
\begin{align}
&-q \, \bm{r} \cdot \bm{E}(t, x) 
= -q \, z \, {\mathcal E}_{\rm{SW}} \, \cos(\omega t) \, \cos(k x) \nonumber\\[2ex]
&\qquad = -q \, z \, 
 {\mathcal E}_{\rm{SW}} \, \cos(\omega t) \,
\left( 1 - \frac{(k x)^2}{2} + {\cal O}(x^4) \right)\,.
\end{align}
This naive expansion gives the wrong prefactor 
and is not applicable here.

The term in Eq.~(\ref{last-term}) entails a spin-independent 
correction to the transition dipole moment. At resonance, one has
\begin{equation}
k = \frac38 \, (Z\alpha)^2 \,m\,.
\end{equation}
The relative correction due to the additional contribution 
Eq.~(\ref{last-term}) to the interaction Hamiltonian
is therefore
\begin{equation}
- \frac{k^2}{6} \, 
\frac{\left< 1S, \! m \! = \! 0 | 
z \, x^2 | 2P, \! m \! = \! 0  \right>}%
{\left< 1S, \! m \! = \! 0 | z | 2P, \! m \! = \! 0 \right>}
=
- \frac{1}{16}\,(Z\alpha)^2\,.
\end{equation}
Analogous to Eq.~(\ref{defdeltadj}), this modification of the
transition dipole moment, for a 1$S$--2$P$ transition,
gives rise to a correction to 
the Rabi frequency given by 
\begin{equation}
\delta \Omega^{(\rm F)} = - \frac{1}{16}\,(Z\alpha)^2 \: \Omega \,.
\end{equation}
The summed shift $\delta \overline{\omega}^{(\rm F)}_{\pm}$ of 
the Mollow sidebands due to this 
modification of the Rabi frequency can be expressed
as
\begin{equation}
\label{summedField}
\delta \overline{\omega}^{(\rm F)}_{\pm} =
\pm \left ( \sqrt{\left (\Omega+
\delta \Omega^{(\rm F)}\right )^2 + \Delta^2}
- \sqrt{\Omega^2 + \Delta^2} \right ) \,,
\end{equation}
where
\begin{equation}
\delta \overline{\omega}^{(\rm F)}_{\pm} 
\approx \delta \omega^{(\rm F)}_{\pm} 
= \mp \frac{\Omega^2}{\sqrt{\Omega^2 + \Delta^2}} \, 
{\mathcal F} 
\end{equation}
and
\begin{equation}
{\mathcal F} = \frac{1}{16}\,(Z\alpha)^2 \, .
\end{equation}
With these definitions, the summed relativistic correction to the dipole
moment effectively corresponds to a replacement
$\Omega \to \Omega (1-{\mathcal F})$.

It is important to note that the long-wave QED correction to the 
interaction Hamiltonian Eq.~(\ref{lwqed}) does not couple any
unwanted magnetic quantum numbers to the laser-driven doublet.

\subsubsection{\label{c-term}Higher--Order Corrections 
(in $\Omega$ and $\Delta$)
to the Self--Energy of Dressed States and Corresponding
Correction to the Rabi Frequency}

The Lamb shift of dressed states is different from the 
Lamb shift of atomic bare states, as already discussed
in Sec.~\ref{intro}. In this section, we extend 
the analysis of Sec.~\ref{Lamb}
to the next-higher order. For this, we keep the terms linear in 
$\Omega$ and $\Delta$ in 
evaluating the energy shifts of the dressed states.
As explained in detail in~\cite{JeEvHaKe2003,JeKe2004}, we
thereby obtain a further correction to the position of the 
fluorescence sidebands which may be expressed as
\begin{equation}
\label{deltaomegaplusminus}
\delta \omega_\pm^{(C)} =
\mp {\cal C}\, \frac{\Omega^2}{\sqrt{\Omega^2 + \Delta^2}}\,,
\end{equation}
where
\begin{eqnarray}
{\cal C} &=& \frac{\alpha}{\pi} \, \ln[(Z\alpha)^{-2}] \,
\frac{\left< \bm{p}^2 \right>_g + \left< \bm{p}^2 \right>_e}{m^2}
\end{eqnarray}
is a dimensionless constant. For the hydrogen 1$S$-2$P$ transition,
the leading logarithmic term is independent of the spin and given by
\begin{eqnarray}
{\cal C} &=& \frac{5}{4\pi} \, \alpha (Z\alpha)^2\, 
    \ln[(Z\alpha)^{-2}] \, \label{corr-c}\, .
\end{eqnarray}
This corrections may be interpreted physically with a summation
as used for the bare Lamb shift correction:
\begin{equation}
\label{summedC}
\delta \overline{\omega}^{(C)}_\pm = \pm
\left( \sqrt{\Omega^2 \, (1 - {\cal C})^2 +
\Delta ^2} - \sqrt{\Omega^2 + \Delta^2}\right)\,,
\end{equation}
with $\delta \overline{\omega}^{(C)}_\pm \approx 
\delta\omega_\pm^{(C)}$ because of the 
smallness of the correction.
Thus the additional
shift $\delta\omega_\pm^{(C)}$ may be interpreted
as a radiative correction $\Omega \to \Omega(1-{\mathcal C})$
of the Rabi frequency.

%
%
\subsubsection{\label{log-dipole}Leading Logarithmic Radiative 
Corrections to the Transition Dipole Moment (Vertex Corrections)}

In ``normal'' bound-state quantum electrodynamics,
vertex corrections are evaluated with respect
to the interaction of the electron with the binding 
field of the atomic nucleus. In an effective treatment,
and in leading logarithmic approximation,
the effect of the self-energy may be accounted for 
by making use of an effective Lamb-shift potential~\cite{Ka1996}
\begin{equation}
\label{Vlamb}
\Delta V_{{\rm Lamb}}({\bm r}) = 
\frac43 \, \alpha (Z\alpha) \ln[(Z\alpha)^{-2}] \,
\frac{\delta^{(3)}({\bm r})}{m^2}\: ,
\end{equation}
which modifies the Coulomb interaction according to 
\begin{equation}
-\frac{Z\alpha}{r} \to
-\frac{Z\alpha}{r}  + \Delta V_{{\rm Lamb}}({\bm r}) \,.
\end{equation}
Note that the potential (\ref{Vlamb}) is really the consequence
of a self-energy (``vertex'') correction, not that of 
vacuum polarization. In many cases, vacuum polarization corrections may
also be accounted for by employing an effective potential,
but the corresponding potential lacks the large 
logarithm $\ln[(Z\alpha)^{-2}]$. An accurate treatment 
of self-energy corrections requires the consideration
of many more terms than the crude approximation (\ref{Vlamb}).

Here, we evaluate the leading vertex corrections
to the interaction of the bound electron with the driving
laser field, and so we have to consider both the 
Coulomb as well as the laser field.
Nonrelativistically, the atom-laser interaction is
given by the matrix element
of the usual interaction Hamiltonian,
which reads (length gauge, $1S \Leftrightarrow 2P$ transition)
$-q E_{\rm L} \langle 1S |z| 2P \rangle = -q E_{\rm L} d$
where $E_{\rm L}$ is the field strength of the 
(strong) laser field and the dipole moment is
\begin{eqnarray}
d &=& \left<2P \left | z \right |1S\right> 
= \frac{2^7}{3^5}\sqrt{2} \frac{1}{m (Z\alpha)} \,.
\end{eqnarray}
Vertex corrections lead to modifications of the dipole moment
given by $d \to d + \delta d$, where
the vertex correction $\delta d$ is considered
below, and the radiative correction (in the length gauge) 
to the laser-atom interaction is effectively
a replacement $- q E_{\rm L}\, d \to - q E_{\rm L}\, (d + \delta d)$. 
The large intensity of the driving laser field is accounted for 
in this formalism because the electric 
laser field strength $E_{\rm L}$
multiplies both the dipole moment matrix element of the 
interaction Hamiltonian
$d$ and the radiative correction
$\delta d$.

Laser photons as well as the spontaneously
emitted photons in the radiative decay 
of excited states are real rather than virtual.
Consequently, the radiative corrections
to the laser-atom interaction on the one hand
and to the radiative decay rate on the other hand 
are related to each other.
In the length gauge, the leading-order expression for the 
spontaneous emission decay rate is
\begin{eqnarray}
\label{def-Gamma}
\Gamma &=&  \frac{4}{3} \, \alpha \, E^3 \, d^2
= \frac{2^8}{3^8} \alpha \, (Z\alpha)^4 \,m \,.
\end{eqnarray}
In order to obtain gauge-invariant results for the 
quantum electrodynamic corrections to $\Gamma$ (while working 
in the length gauge), it is necessary
to consider both radiative vertex corrections to 
the dipole moment $d$ 
and corrections to the (bare) transition frequency 
(energy difference) $E$ (see Refs.~\cite{IvKa1996,Ka1996,SaPaCh2004}). 
In our treatment, the vertex corrections to the dipole moment are 
given in Eqs.~(\ref{deltadlog}), (\ref{deltadnloga}) and (\ref{deltadnlogb}), 
whereas the vertex corrections to the transition frequency enter 
into the radiative correction to the detuning in Eq. (\ref{det-rad}).

In general, the vertex corrections to the laser-atom interaction enter at the
relative order of ${\cal O}(\alpha (Z \alpha)^2 \, \ln[(Z\alpha)^{-2}])$. 
One may wonder why the corrections do not enter
at the relative order  ${\cal O}(\alpha)$. The reason 
is that in interactions with real photons (the square of the
four-momentum being $q^2 = 0$), the 
otherwise dominant correction due to the $F_1$ Dirac form 
factor vanishes, and the remaining terms are then of higher 
order in the $Z\alpha$-expansion.

We now analyze the shift of the resonance frequency
and the shift of the transition dipole moment 
induced by the Lamb-shift potential (\ref{Vlamb}).
The transition energy 
\begin{equation}
E = E_{2P} - E_{1S} =  \frac{3}{8} (Z\alpha)^2  m 
\end{equation}
is shifted by $V_{{\rm Lamb}}$ according to 
\begin{eqnarray}
\delta E^{(\rm log)} &=& - 
\left<1S \left |\Delta V_{{\rm Lamb}}({\bm r}) 
\right |1S\right> \nonumber \\
&=& - \frac{4m}{3\pi}\, \alpha (Z\alpha)^4\, \ln [(Z\alpha)^{-2}]\,,
\end{eqnarray}
because the matrix element of the $2P$ state vanishes.
This yields a relative shift of
\begin{equation}
\frac{\delta E^{(\rm log)} }{E} = - 
\frac{32}{9 \pi} \alpha (Z\alpha)^2\, \ln [(Z\alpha)^{-2}] \, .
\end{equation}
The modification of the matrix element 
due to the corrections to the 1$S$ wave function
amounts to
\begin{align}
\delta& d^{(\rm log)} = 
\left<2P \left |z 
\left (\frac{1}{E-H} \right )^{'} 
\Delta V_{{\rm Lamb}}({\bm r}) \right |1S\right> \nonumber \\
&= \sqrt{2} \frac{2^5\, 
\alpha (Z\alpha)}{3^7 m \pi}\, 
\ln [(Z\alpha)^{-2}]\, \Bigl ( 48 \ln \frac{4}{3} + 131 \Bigr),
\end{align}
where the prime denotes the reduced Green function. Thus the logarithmic
relative correction is~\cite{IvKa1996} 
\begin{equation}
\label{deltadlog}
\frac{\delta d^{(\rm log)} }{d} = 
\frac{\alpha (Z\alpha)^2}{\pi} \, 
\ln [(Z\alpha)^{-2}]\, 
\left ( \frac{4}{3}\, \ln \frac{4}{3} + \frac{131}{36} \right) \, .
\end{equation}
This leads to a correction to the Rabi frequency 
analogous to Eq.~(\ref{relom}),
\begin{equation}
\frac{\delta \Omega^{(\rm log)}}{\Omega} = 
\frac{\delta d^{(\rm log)} }{d}\,,
\end{equation}
where we may ignore the spin, in contrast to (\ref{deltarREL}).

The interpretation of this shift is analogous to the relativistic corrections
to the dipole matrix element in Sec.~\ref{rel-dipole}. The Mollow sidebands
are shifted by the frequency
\begin{equation}
\delta \overline{\omega}^{(\rm log)}_\pm = \pm \left ( \sqrt{(\Omega + \delta \Omega^{(\rm log)} )^2 + \Delta^2} 
     - \sqrt{\Omega^2 + \Delta^2} \right ) 
\end{equation}
where     
\begin{equation}
\delta \overline{\omega}^{(\rm log)}_\pm \approx
\delta \omega^{(\rm log)}_\pm 
= \pm \frac{\Omega^2}{\sqrt{\Omega^2 + \Delta^2}} \, {\mathcal A} \, ,
\end{equation}
and
\begin{eqnarray}
{\mathcal A} &=& \frac{\delta d^{(\rm log)}}{d} + {\mathcal O}\left ( \frac{\delta d^{(\rm log)}}{d} \right )^2 
\end{eqnarray}
is a dimensionless constant. Then these corrections may be accounted 
for by the replacement $\Omega \to \Omega (1+{\mathcal A})$.
One should note that at this parametric order the above results
hold for both states 2$P_{\nicefrac{1}{2}}$ and 
2$P_{\nicefrac{3}{2}}$~\cite{SaPaCh2004,IvKa1996}.

%
%
\subsubsection{\label{nlog}Nonlogarithmic Vertex and Vacuum 
Polarization Corrections\\ to the Transition Dipole Moment}

The concurrence of the radiative shifts to the 
transition dipole matrix elements for the  
1$S$---2$P_{\nicefrac{1}{2}}$ and the 
1$S$---2$P_{\nicefrac{3}{2}}$ transitions,
which was found in Sec.~\ref{log-dipole}
for the effects of relative order $\alpha (Z\alpha)^2\,\ln[(Z\alpha)^{-2}]$,
is lifted on taking into account 
corrections of order $\alpha (Z\alpha)^2$ 
(no logarithms) due to the self-energy.
Vacuum polarization corrections 
to the transition dipole matrix elements also 
enter at the relative order of $\alpha\,(Z\alpha)^2$. 

The sum of the nonlogarithmic (``nlog'')
vertex and vacuum polarization corrections
to the transition dipole moments 
of order $\alpha\,(Z\alpha)^2$ are given by~\cite{SaPaCh2004}
\begin{subequations}
\label{deltadnlog}
\begin{equation}
\label{deltadnloga}
\frac{\delta d_{\nicefrac{1}{2}}^{(\rm nlog)}}{d_{\nicefrac{1}{2}}} = 
    - \frac{\alpha (Z\alpha)^2}{\pi} \: 9.2(1.8)
\end{equation}
for the 2$P_{\nicefrac{1}{2}}$ state and by
\begin{equation}
\label{deltadnlogb}
\frac{\delta d_{\nicefrac{3}{2}}^{(\rm nlog)}}{d_{\nicefrac{3}{2}}} =
    - \frac{\alpha (Z\alpha)^2 }{\pi}\: 9.3(1.9)
\end{equation}
\end{subequations}
for the 2$P_{\nicefrac{3}{2}}$ state~\cite{SaPaCh2004}.
In between Eqs.~(70) and~(71) of~\cite{SaPaCh2004}, 
it is stated that currently, there is an internal discrepancy 
between the numerically obtained values for the radiative 
correction at $Z=5, 10, \dots$
on the one hand and analytic results for the first terms of
the $Z\alpha$-expansion (logarithm$+$constant)
on the other hand. This discrepancy is 
of the order of $10\,\%$ of the total constant
term of order $\alpha(Z\alpha)^2$, and this 
limits the current status of the theory.
Here, we employ an even more conservative error estimate 
and assign a $20\,\%$ uncertainty
to both of the numerical values in Eqs.~(\ref{deltadnloga}) 
and~(\ref{deltadnlogb}).
The corresponding correction to the Rabi frequency is 
again analogous to Eq.~({\ref{relom}),
\begin{equation}
\frac{\delta \Omega^{(\rm nlog)}_j}{\Omega} = 
\frac{\delta d^{(\rm nlog)}_j }{d_j} \, .
\end{equation}
One obtains the following total 
correction $\delta \overline{\omega}^{(\rm TDM)}_{\pm,j}$
to the position of the sidebands due to logarithmic and nonlogarithmic
correction to the transition dipole matrix element (TDM):
\begin{align}
\label{summedTDM}
 \delta \overline{\omega}^{(\rm TDM)}_{\pm,j} 
=& \pm \! \left ( \sqrt{\left (\Omega 
+ \delta \Omega^{(\rm log)} + 
\delta \Omega_j^{(\rm nlog)} \right )^2 + \Delta^2}  \right. \nonumber \\
&\left. - \sqrt{\Omega^2 + \Delta^2} \right ) 
\end{align}
where 
\begin{equation}
\delta \overline{\omega}^{(\rm TDM)}_{\pm,j} \approx
\delta \omega^{(\rm TDM)}_{\pm,j} = 
\pm \frac{\Omega^2}{\sqrt{\Omega^2 + \Delta^2}} \, {\mathcal A}_j
\end{equation}
and
\begin{eqnarray}
{\mathcal A}_j &=& \frac{\delta d^{\rm (log)}}{d} + 
\frac{\delta d_j^{\rm (nlog)}}{d_j} + 
{\mathcal O}\left ( \alpha (Z\alpha)^3 \right )\,.
\end{eqnarray}
${\mathcal A}_j$
is a dimensionless constant and $j=\nicefrac{1}{2}, \nicefrac{3}{2}$. 
Thus the corrections amount to a modification of the Rabi frequency
given by $\Omega \to \Omega (1+{\mathcal A}_j)$.

We expect a similar nonlogarithmic 
correction of order $\alpha(Z\alpha)^2$ to supplement
the $\mathcal C$-term discussed in Sec.~\ref{c-term}.
The logarithmic $\mathcal C$-term was found to be of the order of 
$\alpha(Z\alpha)^2\,\ln[(Z\alpha)^{-2}]$.
The evaluation of the corresponding nonlogarithmic term 
is however beyond the scope of this work. Here, we only
present a (conservative) estimate of the expected correction.
To this end we observe that the general form of the 
radiative corrections under discussion is 
\begin{equation}
\alpha(Z\alpha)^2\: 
\ln [(Z\alpha)^{-2}] \cdot c_1 + \alpha(Z\alpha)^2\cdot c_2\, ,
\end{equation}
where $c_1$ and $c_2$ are dimensionless constants. 
Based on experience with similar 
corrections (both for self-energy effects and well as
radiative corrections to decay rates~\cite{SaYe1990,SaPaCh2004}), 
we assume the following 
relation with the corresponding uncertainty
for the unknown parameter $c_2$:
\begin{equation}
c_2 = (-2 \pm 2) \cdot c_1\, . \label{c-estimate}
\end{equation}
For example, we verify the validity
of this estimate for the particular contribution
to the nonlogarithmic part discussed above, 
i.e.~for the terms that contribute to ${\mathcal A}_j$.
For these terms, the estimate $(-2 \pm 2) \cdot c_1$
evaluates to $-2.56 \pm 2.56$ for the 
constant term. In comparison,
the values obtained above for $c_2$ by a direct numerical
analysis are $-2.92$ for the correction
${\mathcal A}_{\nicefrac{1}{2}}$,
and $-2.97$ for the correction
${\mathcal A}_{\nicefrac{3}{2}}$, which agrees to
the estimate.
For the ${\mathcal C}$-term correction, we thus obtain from
Eqs.~(\ref{corr-c}) and (\ref{c-estimate})
\begin{equation}
{\mathcal C}_j = \alpha (Z\alpha)^2 \left ( \frac{5}{4\pi}\:\ln [(Z\alpha)^{-2}]
 - \frac{5}{4\pi} (2\pm 2)  \right )  
\end{equation}
as the combination of the logarithmic and the estimated nonlogarithmic
correction. We expect the nonlogarithmic correction to be 
spin-dependent in analogy to (\ref{deltadnlog}).
As for the correction to the matrix element, 
this shifts the Mollow sidebands by [see also~Eq.~(\ref{summedC})]
\begin{eqnarray}
\label{summedCj}
\delta \overline{\omega}^{(C)}_{\pm,j} 
&= &\pm \left ( \sqrt{\Omega^2(1-{\mathcal C}_j)^2  + \Delta^2} 
     - \sqrt{\Omega^2 + \Delta^2} \right ) \nonumber \\
&\approx& \mp \frac{\Omega^2}{\sqrt{\Omega^2 + \Delta^2}} \, {\mathcal C}_j \, ,
 \end{eqnarray}
such that the correction may be applied by the replacement 
$\Omega \to \Omega (1-{\mathcal C}_j)$.

%
%
\subsubsection{\label{sec-secular}Corrections to the Secular Approximation}

In this section, we transform the correction terms
to the secular approximation in Eq.~(\ref{secular-corrections}) such that 
they may be integrated into our correction scheme.
Eq.~(\ref{secular-corrections}) may be rewritten as
\begin{eqnarray}
\omega_\pm &=& \omega_{\rm L} \pm \Omega_{\rm R} 
\mp \frac{\Omega^2}{\sqrt{\Omega^2+\Delta^2}}\, {\mathcal S} \, ,\\
{\mathcal S} &=& \frac{4+y^2}{8(1+y^2)}\, \left ( \frac{\Gamma}{\Omega}
\right )^2 + {\mathcal O} \left ( \frac{\Gamma}{\Omega} \right )^4 \, . 
\label{cor-s}
\end{eqnarray}
For $\Delta \ll \Omega$, one may expand the leading contribution
to ${\mathcal S}$ to give
\begin{eqnarray}
{\mathcal S} = \frac{1}{2}\,\left ( \frac{\Gamma}{\Omega} \right )^2
+ {\mathcal O} \left ( \frac{\Delta^2\Gamma^2}{\Omega^4} \right ) \, . \label{par-s}
\end{eqnarray}
With this definition, the corrections to the secular approximation 
may be accounted for in the final result with the replacement
$\Omega \to \Omega \, (1 - {\mathcal S})$, which results in
a summed shift of
\begin{equation}
\label{summedS}
\delta \overline{\omega}^{(S)}_{\pm} 
= \pm \left ( \sqrt{\Omega^2(1 - {\mathcal S})^2  + \Delta^2} 
- \sqrt{\Omega^2 + \Delta^2} \right ) \,.
\end{equation}

\section{\label{num-analysis}
NUMERICAL DATA FOR THE HYDROGEN $1S$--$2P_j$ TRANSITIONS
($j = \nicefrac{1}{2}, \nicefrac{3}{2}$)}

In the previous section, we have discussed both 
corrections to the detuning and to the Rabi frequency, which give
rise to a modification of the positions
of the sidebands in the Mollow spectrum. 
We start the evaluation of correction terms
here from a point where we assume that
all relativistic corrections to the 
transition frequency, as well as hyperfine-structure
effects, have already been included in the 
bare transition frequency. This frequency corresponds to
the prediction obtained using Eq.~(\ref{defElamb})
by setting $L_{nl_j}$ explicitly equal to zero.
We therefore redefine the detuning $\Delta$ to be 
\begin{equation}
\label{redefine}
\Delta = \omega_{\rm L} - \tilde{\omega}_{\rm R}\,,
\end{equation}
where [cf.~Eq.~(\ref{defElamb})]
\begin{equation}
\tilde{\omega}_{\rm R} = \tilde{E}(2P_j) - \tilde{E}(1S)
\end{equation}
with
\begin{eqnarray}
\label{deftildeElamb}
\lefteqn{\tilde{E}(nl_j) = m_r \left[ f(n,j)-1 \right]} \nonumber\\
& & - \frac{m_r^2}{2 (m + m_N)}
\left[ f(n,j) - 1 \right]^2 + E_{\rm hfs}\,.
\end{eqnarray}
Of course, the 
full theoretical prediction $E(nl_j)$ is obtained 
as the sum of $\tilde{E}(nl_j)$ and $L_{nl_j}$.
The modification of the Rabi frequency
due to the bare Lamb shift 
\begin{equation}
L^{(j)}_{\rm bare} = L_{2P_j} - L_{1S}\,,
\end{equation}
as discussed in Sec.~\ref{Lamb},
finds a natural interpretation as a contribution to the 
{\em Lamb shift of the dressed states}.
The various first-order 
correction terms to the lowest-order prediction
for the generalized Mollow-sideband displacement
\begin{equation}
\pm \sqrt{\Omega^2 + \Delta^2}\,,
\end{equation}
starting from (\ref{redefine}),
may be summarized as follows:
\begin{eqnarray}
\label{total1}
&&\delta \omega_\pm^{(j)} = \mp \frac{\Delta}{\sqrt{\Omega^2 + \Delta^2}}
\left( L_{\rm bare}^{(j)} + 
{\mathcal B}\, \Omega^2 + {\mathcal D}_{\rm R}\, \Omega^2\right) 
\nonumber \\
&& \pm  \frac{\Omega^2}{\sqrt{\Omega^2 + \Delta^2}} 
\left(  {\mathcal A}_j  - {\mathcal C}_j - 
{\mathcal E}_j - {\mathcal F} - {\mathcal S} \right) \, ,
\label{firstorder}
\end{eqnarray}
where 
\begin{subequations}
\label{params}
\begin{align}
L_{\rm bare}^{(\nicefrac{1}{2})} =& 
-8185.647(34) \cdot 10^6 \, {\rm Hz} \, , \\[1ex]
L_{\rm bare}^{(\nicefrac{3}{2})} =& 
-8160.294(34) \cdot 10^6 \, {\rm Hz} \, , 
\displaybreak[0] \\[1ex]
{\mathcal B} =& \frac{1}{4 \omega_{\rm R}}  \, ,\\[1ex]
{\mathcal D} =&  {\mathcal D}_{\rm R} + 
{\rm i} \:{\mathcal D}_{\rm I} \nonumber \\[1ex]
=& \frac{1}{\omega_{\rm R}} \,
\left[ 2.3305(3) - 0.088245(6)\, {\rm i} \right],
\displaybreak[0]  \\[1ex]
{\mathcal A} =& 
\frac{\alpha (Z\alpha)^2}{\pi} 
\ln(Z\alpha)^{-2} \left ( \frac{4}{3} \ln \frac{4}{3} \! + \!
\frac{131}{36} \right) , \\[1ex]
{\mathcal A}_{\nicefrac{1}{2}} =& 
{\mathcal A} -\frac{\alpha(Z\alpha)^2}{\pi}\: 9.2(1.8) \, ,\\[1ex]
{\mathcal A}_{\nicefrac{3}{2}} =& 
{\mathcal A} -\frac{\alpha(Z\alpha)^2}{\pi}\: 9.3(1.9) \, ,
\displaybreak[0] \\[1ex]
{\mathcal C}_j =& 
\alpha (Z\alpha)^2 \frac{5}{4\pi} \left ( \ln [(Z\alpha)^{-2}]
 - (2\pm 2) \right )\,,  \\[1ex]
{\mathcal E_{\nicefrac{1}{2}}} =& 
(Z\alpha)^2 \left(\frac{13}{32}+\frac{3}{2}\ln 2 - \ln 3 \right) \, ,\\[1ex]
{\mathcal E_{\nicefrac{3}{2}}} =& 
(Z\alpha)^2\left(\frac{31}{96}+\frac{5}{4}\ln 2 - 
\frac{3}{4} \ln 3 \right) \, , \\
{\mathcal F} =& \frac{1}{16}\,(Z\alpha)^2 , \\
{\mathcal S} =& \frac{1}{2} \left ( \frac{\Gamma}{\Omega} \right )^2 \, .
\end{align}
\end{subequations}
We have found
that the bare Lamb shift $L_{\rm bare}^{(j)}$, 
the Bloch-Siegert shift (${\mathcal B}$), and 
the off-resonant self-energy corrections (${\mathcal D}_{\rm R}$) 
give rise
to a modification of the detuning $\Delta$ in the expression
for the Mollow spectrum. 
As the latter two effects are 
intensity-dependent, also their correction to the detuning 
depends on the intensity of the incident laser field.
The modifications to the
transition dipole moment (${\mathcal A}_j, {\mathcal E}_j, {\mathcal F}$)
 as well as the higher-order resonant
self-energy shifts (${\mathcal C}_j$) and the correction to the 
secular approximation ($\mathcal S$)
may be interpreted as radiative corrections
to the Rabi frequency $\Omega$.
The interpretations as a modification of the detuning and
the Rabi frequency may best be seen by using a summation of
Eq.~(\ref{total1}), which we have shown to be valid up to first 
order in the parameters in Eq.~(\ref{params}):
\begin{equation}
\label{fullsummation}
\delta \overline{\omega}_\pm^{(j)} = 
\pm \left( \Omega^{(j)}_{\mathcal C} 
- \sqrt{\Omega^2 + \Delta^2}\right)
\end{equation}
with
\begin{equation}
\label{fullOmega}
\Omega^{(j)}_{\mathcal C} =
\sqrt{\Omega^2 \cdot \left(1+{\hat{\Omega}_{\rm rad}^{(j)}}\right)^2 +
\left(\Delta - {\Delta^{(j)}_{\rm rad}}\right)^2}\,.
\end{equation}
Here, the Rabi frequency and detuning are supplemented by the
discussed relativistic and radiative corrections;
these are given by 
\begin{subequations}
\begin{eqnarray}
\Delta^{(j)}_{\rm rad} &=&  L^{(j)}_{\rm bare} + 
{\mathcal B}\, \Omega^2 + {\mathcal D}_{\rm R}\, 
\Omega^2 \, , \label{det-rad}\\
\hat{\Omega}_{\rm rad}^{(j)} &=&   
{\mathcal A}_j - {\mathcal C}_j - {\mathcal E}_j  - 
{\mathcal F} -{\mathcal S}\, . \label{rabi-rad}
\end{eqnarray}
\end{subequations}
This summation implied by (\ref{fullsummation})
is motivated by Eqs.~(\ref{summedREL}),
(\ref{summedLamb}), (\ref{summedBS}), (\ref{summedOR}),
(\ref{summedR}), (\ref{summedC}), (\ref{summedTDM}), 
(\ref{summedField}), (\ref{summedCj}), and (\ref{summedS}).
In (\ref{rabi-rad}), the symbol $\hat{\Omega}_{\rm rad}^{(j)}$ indicates
a relative modification of the Rabi frequency,
i.e.~a dimensionless quantity.

We have thus ``summed'' all the radiative corrections 
as effective corrections to the Rabi frequency
and the detuning.  Of course, the mixing angle $\theta$ as
defined in Eqs.~(\ref{defmixing}) and (\ref{matchmixing})
is changed by the radiative
corrections. Indeed, one may evaluate the corrected
$\theta$ by employing the relation
$\tan(2 \theta) = - \Omega_{\rm corr}/\Delta_{\rm corr}$
where $\Delta_{\rm corr} = \Delta - {\Delta^{(j)}_{\rm rad}}$ and
$\Omega_{\rm corr} = 
\Omega \cdot \left(1+{\hat{\Omega}_{\rm rad}^{(j)}}\right)$ are
the relativistically and radiatively corrected Rabi frequency and
the detuning, respectively [see Eqs.~(\ref{rabi-rad}) 
and~(\ref{det-rad})]. Because all relativistic and radiative
corrections find a natural interpretation
as corrections to the Rabi frequency and the detuning,
the corrected dressed states have the same
structure as Eqs.~(\ref{plus}) and (\ref{minus}),
but with relativistic and radiative wave function corrections 
a corrected mixing angle.

All relativistic and radiative
corrections to the Rabi frequency and the detuning
have been evaluated here using the unperturbed mixing angle
$\theta$. The ``corrections to the corrections''
(sic!) due to an evaluation of modifications to the Rabi frequency
and the detuning in terms of the corrected mixing angle
are of higher order than the terms relevant
for the discussion in the current paper and may be neglected
on the level of approximation employed in the current 
investigation.}

In the following numerical analysis, we assume the atom to be located
at an anti-node of a laser field in standing-wave configuration.
The atom is thus driven by two counterpropagating laser beams,
whereas in the definitions of the electric field and the Rabi
frequency in Sec.~\ref{results} and especially in the matching of
the classical macroscopic field  with the corresponding quantum counterpart 
in Eq.~(\ref{matching}) a single-mode running-wave field was 
considered. However all results of Sec.~\ref{calculation} also apply to a 
standing wave field configuration if the macroscopic electric field strength 
${\mathcal E}_{\rm L}$ is taken to be the total field strength of both 
counterpropagating field modes at the position of the atom. In order to
avoid confusions, from now on we thus denote the total electric field strength
of the standing wave as ${\mathcal E}_{{\rm SW}}$.

%
%
\subsection{\label{num-1s2p12}$1S_{\nicefrac{1}{2}} 
\leftrightarrow 2P_{\nicefrac{1}{2}}$}

For the $|2P_{\nicefrac{1}{2}}, m\!=\!\pm\nicefrac{1}{2}\rangle$ state
as upper state, the decay constant is given by 
$\Gamma_{\nicefrac{1}{2}} = 99.70942(1)\cdot 10^6 $ Hz \cite{SaPaCh2004}.
In order to account for the dependence on the
laser field intensity, we introduce the
parameter $h_{\nicefrac{1}{2}}=|\Omega|/\Gamma_{\nicefrac{1}{2}}$. Then 
for $h_{\nicefrac{1}{2}}=1000$ and
$\Delta=50\:\Gamma_{\nicefrac{1}{2}}$, one has 
$\omega_{\rm R}~\gg~\Omega~\gg~\Gamma_{\nicefrac{1}{2}}, \Delta$.
Therefore,
the relative corrections to the detuning and the Rabi frequency 
in Eqs.~(\ref{det-rad}) and (\ref{rabi-rad}) become
\begin{subequations}
\begin{eqnarray}
\Delta^{(\nicefrac{1}{2})}_{\rm rad} &=&  
-8.175249(33) \cdot 10^9  {\rm Hz} \, ,\\*
\Omega^{(\nicefrac{1}{2})}_{\rm rad} &=&  
-19.78(56) \cdot 10^{-6} \,.
\end{eqnarray}
\end{subequations}
Here, the parameter $h_{\nicefrac{1}{2}}$  may be expressed 
in terms of the electric field strength ${\cal E}_{\rm L}$
as
\begin{equation}
h_{\nicefrac{1}{2}} = 346.783 \cdot  10^{-6} \: 
\left |{\cal E}_{{\rm SW}}\left[{\rm V}/m \right] \right | \: .     
\end{equation}
Of course, $\left[{\rm V}/m \right]$ in this case means that the 
peak electric field strength of the laser is assumed to be measured
in Volts per meter.

The absolute ionization rate $\mathcal I$ into 
the continuum due to the driving laser field
is given by ${\mathcal I} = {\mathcal D}_{\rm I} \: \Omega^2$.
In an experiment, this ionization rate has to be 
much smaller than the Rabi frequency, such that on average
the atom undergoes many fluorescence cycles before it is ionized. 
Thus we define the ratio 
\begin{equation}
{\mathcal I}_\Omega = 
\left| \frac{\mathcal I}{\Omega}\right| = 
\left|{\mathcal D}_{\rm I}\right| \: \Omega = 
\left| {\mathcal D}_{\rm I}\right| 
\: h_{\nicefrac{1}{2}} \: \Gamma_{\nicefrac{1}{2}} \, ,
\end{equation}
which has to be much smaller than unity [${\mathcal I}_\Omega \ll 1$].
For $h_{\nicefrac{1}{2}}=1000$, one obtains 
\begin{eqnarray}
{\mathcal I} (h_{\nicefrac{1}{2}}=1000) &=& 356 \: \textrm{kHz}\, , \\
{\mathcal I}_\Omega (h_{\nicefrac{1}{2}}=1000) &=& 3.6\cdot 10^{-6}\, ,
\end{eqnarray}
which means that the probability
of one-photon ionization does not restrict the above measurement scheme.

\begin{table}[t]
\begin{ruledtabular}
\begin{tabular}{|c|.|.|}
\hline 
\multicolumn{1}{|c|}{\rule[-3mm]{0mm}{8mm} 
Shift} & 
\multicolumn{1}{c|}{1$S_{\nicefrac{1}{2}} 
\leftrightarrow$ 2$P_{\nicefrac{1}{2}}$ [kHz]} & 
\multicolumn{1}{c|}{1$S_{\nicefrac{1}{2}} 
\leftrightarrow$ 2$P_{\nicefrac{3}{2}}$ [kHz]} \\ 
\hline
\hline 
\rule[-3mm]{0mm}{8mm} 
$\delta \overline{\omega}^{\rm (Lamb)}_{+,j}$ 
& 741599x(4)      & 738281x(4)  \\ 
\rule[-3mm]{0mm}{8mm} 
$\delta \overline{\omega}^{\rm (BS)}_+$ 
& -50x.30(5)   & -50x.30(5) \\ 
\rule[-3mm]{0mm}{8mm} 
$\delta \overline{\omega}^{\rm (OR)}_+$ 
& -468x.51(6)   & -468x.51(6) \\ 
\hline
\hline 
\rule[-3mm]{0mm}{8mm} 
$\delta \overline{\omega}^{\rm (R)}_{+,j}$ 
& -1842x.1(1)    & -1937x.7(1) \\ 
\rule[-3mm]{0mm}{8mm} 
$\delta \overline{\omega}^{\rm (F)}_{+}$ 
& -331x.44(2)    & -331x.44(2) \\ 
\rule[-3mm]{0mm}{8mm} 
$\delta \overline{\omega}^{\rm (C)}_{+,j}$ 
& -121x(31)     &  -121x(31) \\
\rule[-3mm]{0mm}{8mm} 
$\delta \overline{\omega}^{\rm (TDM)}_{+,j}$ 
& 374x(25)    & 372x(26)   \\ 
\rule[-3mm]{0mm}{8mm} 
$\delta \overline{\omega}^{(S)}_{+}$ & -49x.8(2)   & -49x.8(2)  \\ 
\hline 
\end{tabular}
\caption{\label{tabular}Summary of all individual energy shifts due to the 
various discussed corrections. All numbers are obtained for
$h_j=1000$ and $\Delta = 50 \cdot \Gamma_j$ 
with $j=\nicefrac{1}{2}$ ($j=\nicefrac{3}{2}$) 
for the left (right) column using
summation formulas such as Eq.~(\ref{fullsummation}).
Here, $\delta \overline{\omega}^{\rm (Lamb)}_+$ 
is the correction to the high-frequency Mollow sideband position 
related to the bare Lamb shift (c.f. Sec. \ref{Lamb}), 
the symbol (BS) denotes Bloch-Siegert shifts (Sec. \ref{bloch}), 
and the (OR)-shifts are 
due to off-resonant excitations (Sec. \ref{sec-offres}). These shifts 
all may be interpreted as arising from a modified detuning $\Delta$
and are discussed in Sec.~\ref{sCorrDetuning}. 
The other five corrections
are due to a modified Rabi frequency (Sec.~\ref{sCorrRabi}). In particular, 
$\delta \overline{\omega}^{\rm (R)}_+$ is discussed in
Sec. \ref{rel-dipole} and refers to relativistic corrections,
whereas $\delta \overline{\omega}^{\rm (F)}_+$ (Sec. \ref{field-config})
is a field-configuration dependent shift.
The shift $\delta \overline{\omega}^{\rm (TDM)}_+$ 
(Secs. \ref{log-dipole} and  \ref{nlog}) refers to radiative corrections 
to the transition dipole matrix element, and
$\delta \overline{\omega}^{\rm (C)}_+$ 
is a dynamic correction to the Rabi frequency (Secs. \ref{c-term}
and \ref{nlog}). Finally, $\delta \overline{\omega}^{(S)}_{+}$ is a shift due to corrections 
to the secular approximation 
(Sec.~\ref{sec-secular}).}
\end{ruledtabular}
\end{table}
%
%
%

For $h_{\nicefrac{1}{2}}=1000$ and 
$\Delta=50\:\Gamma_{\nicefrac{1}{2}}$, the theoretical
prediction for the shift of the Mollow sidebands 
relative to the central Mollow peak 
by the generalized corrected Rabi frequency 
is as follows:
\begin{equation}
\label{pred12}
\pm \Omega^{(\nicefrac{1}{2})}_{\mathcal C} = 
\pm \: 100.572258(60) \cdot 10^{9} \: {\rm Hz}\,.
\end{equation}
This formula has been evaluated using the
summation formula Eq.~(\ref{fullsummation}) and 
includes all corrections,
in particular the ${\mathcal C}$-term evaluated in
Sec.~\ref{c-term}. For comparison, we also give here
a theoretical prediction that would be obtained
by ignoring the ${\mathcal C}$-term,
\begin{equation}
\label{om12ohneC}
\pm \Omega^{(\nicefrac{1}{2})}_{\rm no\,{\mathcal C}} =
\pm \: 100.572377(27) \cdot 10^{9} \: {\rm Hz} \, .
\end{equation}
This result is obtained by explicitly
setting ${\mathcal C}_j$ in Eq.~(\ref{rabi-rad}) equal
to zero, but still using the full summation according to Eq.~(\ref{fullOmega})
for all other corrections. 
A comparison of Eq.~(\ref{pred12}) 
to Eq.~(\ref{om12ohneC}) shows that the theoretical uncertainties 
of the two results do not overlap. Therefore, the current
status of the theory would allow to discern the presence 
or absence of the $\mathcal{C}$-term corrections by means
of an accurate experiment.

The principal uncertainty of the generalized Rabi 
frequency in Eq.~(\ref{pred12}) is due to the
uncertainty in the estimate of the nonlogarithmic contribution to 
${\mathcal C}_j$ and ${\mathcal A}_j$. As a reference, the bare
shift without corrections is given by
\begin{equation}
\pm \sqrt{\Omega^2 + \Delta^2} = \pm \:  99.833975 \cdot 10^{9} \: {\rm Hz} \, .
\end{equation}
This result has to be compared to the radiatively corrected
result (\ref{pred12}).

In Table~\ref{tabular}, the individual shifts due to the considered
corrections are listed together with their respective uncertainties
for both transitions 
$1S_{\nicefrac{1}{2}}\leftrightarrow 2P_{\nicefrac{1}{2}}$ and 
$1S_{\nicefrac{1}{2}}\leftrightarrow 2P_{\nicefrac{3}{2}}$.
All shifts are evaluated in their 
``summed'' form [see e.g.~Eq.~(\ref{summedLamb})].
For the bare Lamb shift corrections 
$L_{\rm bare}^{(j)}$, the uncertainty in the shift is mainly due to the
uncertainty in the numerical value for the Lamb shift of the
hydrogen $1S$ state, see Eq.~(\ref{Lamb-1s}).
The Bloch-Siegert shift acquires a numerical uncertainty
due to neglected
terms of relative order ${\mathcal O}(\Omega/\omega_{\rm L})$ from the 
higher-order Bloch-Siegert-type shifts 
(higher-order perturbation theory in the 
counter-rotating terms).
There is a further source of uncertainty
for the Bloch-Siegert shifts due to terms of relative order 
${\mathcal O}(\Delta/\omega_{\rm L},\Omega/\omega_{\rm L})$
from the expansion leading from Eq.~(\ref{shift-b}) to Eq.~(\ref{par-b}).
The shift due to off-resonant excitation $\mathcal D$ has 
an  uncertainty because contributions to the energies 
Eq.~(\ref{energies-d}) of relative order $\Delta/\omega_{\rm R}$
and of order $\Omega/\omega_{\rm R}$
have been neglected in evaluating the $t$-parameters in Eq.~(\ref{t-d}).
There are also  uncertainties
of the Rabi-frequency shifts due to relativistic corrections to the 
transition dipole matrix elements (${\mathcal E}_j$), which 
are due to neglected higher-order corrections of relative order 
$(Z\alpha)^2$ with respect to the leading corrections.
The field-dependent corrections have an uncertainty due to 
higher-order effects of relative order $(Z\alpha)^2$.
The main uncertainty of the $\mathcal C$-term and ${\mathcal A}_j$-term
corrections are due to the uncertainty which we assign to
the nonlogarithmic contribution of relative
parametric order $\alpha(Z\alpha)^2$
[see Eqs.~(\ref{deltadnlog}),(\ref{c-estimate})]. 
The main uncertainty of the shift due to corrections 
to the secular approximation ($\mathcal S$) are due to higher-order terms
of the expansion leading from Eq.~(\ref{cor-s}) to 
Eq.~(\ref{par-s}) or relative order
${\mathcal O}(\Delta^2/\Omega^2)$, and of fourth-order corrections 
to the secular approximation of relative
order ${\mathcal O}(\Gamma^2/\Omega^2)$.
For the entries of Table~\ref{tabular}, the 
shifts have been evaluated individually
according to Eqs.~(\ref{summedREL}),
(\ref{summedLamb}), (\ref{summedBS}), (\ref{summedOR}),
(\ref{summedR}), (\ref{summedField}), (\ref{summedC}), (\ref{summedTDM}),
(\ref{summedCj}), and (\ref{summedS}).

It is perhaps worthwhile to note that one cannot
simply add the corrections to the quasi-energy of dressed states
in the same sense as corrections to the energy of bare atomic 
states. For the evaluation of a theoretical Lamb-shift prediction
of a bare atomic state, the usual procedure is 
to list the various corrections and to  
simply add these in order to give a theoretical prediction for the 
total energy shift (see e.g.~Tabs.~I and II of~\cite{JePa1996}).
For the Lamb shift of laser-dressed states, the natural interpretation
of the corrections implies modifications of the decisive 
physical parameters that determine the quasi-energy of the 
dressed levels: these are the Rabi frequency $\Omega$ and the detuning
$\Delta$. This interpretation implies, however, summations
of the expressions which agree well with the 
first-order results, so that at least for numerically
small corrections, it is still permissible to 
simply add the correction terms in an approximative sense.
The correspondence of first-order expressions and summed 
results holds approximately unless the correction is large compared
to $\Omega$ and $\Delta$; in this case the 
summation yields a different result as compared to the 
first-order expression. This is the case for 
the numerically dominant effect referred to as 
$\delta \overline{\omega}^{\rm (Lamb)}_{+,j}$ in Tab.~\ref{tabular}.
As already discussed in Secs.~\ref{Lamb} and~\ref{unified},
this summation is somewhat nontrivial in particular for the Lamb shift
corrections. Because fundamental symmetry
properties prevent the radiative corrections from 
coupling $S$ and $P$ states (even in higher order), 
preference is given to the summed results.
In principle, more explicit
higher-order calculations outlined in Sec.~\ref{unified}
would be desirable to verify the summations for all
radiative effects discussed here.

\subsection{\label{num-1s2p32}$1S_{\nicefrac{1}{2}} 
\leftrightarrow 2P_{\nicefrac{3}{2}}$}

In this section, we repeat the above numerical analysis
with $|1S_{\nicefrac{1}{2}}, m\!\!=
\!\!\pm \nicefrac{1}{2}\rangle$ as the ground state and 
$|2P_{\nicefrac{3}{2}}, m\!\!=\!\!\pm \nicefrac{1}{2}\rangle$ 
as the excited state.
The decay width of the $2P_{\nicefrac{3}{2}}$ state is given by 
$\Gamma_{\nicefrac{3}{2}} = 99.70942(1) \cdot 10^6 $ Hz  \cite{SaPaCh2004}.
In order to account for the dependence on the
laser field intensity, we introduce the
parameter $h_{\nicefrac{3}{2}}=|\Omega|/\Gamma_{\nicefrac{3}{2}}$. Then 
for $h_{\nicefrac{3}{2}}=1000$ and $\Delta = 50 \Gamma_{\nicefrac{3}{2}}$,
one has $\omega_{\rm R} \gg \Omega \gg \Gamma_{\nicefrac{3}{2}}, 
\Delta$ such that
the corrections to the detuning and the Rabi frequency 
in Eqs.~(\ref{det-rad}) and (\ref{rabi-rad})
become
\begin{subequations}
\begin{eqnarray}
\Delta^{(\nicefrac{3}{2})}_{\rm rad} &=&  
-8.149896(33)  \cdot 10^9 {\rm Hz} \, ,\\
\Omega^{(\nicefrac{3}{2})}_{\rm rad} &=& 
-20.76(56) \cdot 10^{-6} \, .
\end{eqnarray}
\end{subequations}
Here, the parameter $h_{\nicefrac{3}{2}}$ may be expressed 
in terms of the electric field strength ${\cal E}_{\rm L}$
as
\begin{equation}
h_{\nicefrac{3}{2}} = 
490.425 \cdot 10^{-6}\: 
\left |{\cal E}_{{\rm SW}}\left[{\rm V}/m \right] \right | \: .
\end{equation}
For this transition, the ionization to the continuum is again given by
\begin{eqnarray}
{\mathcal I} (h_{\nicefrac{3}{2}}=1000) &=& 356 \: \textrm{kHz}\, , \\
{\mathcal I}_\Omega (h_{\nicefrac{3}{2}}=1000) &=& 3.6\cdot 10^{-6}\, .
\end{eqnarray}
Thus the scheme is not restricted by ionization on this transition
either.

With the above parameters for the Rabi frequency and the 
detuning, the positions of the Mollow sidebands
relative to the center component with full corrections, without
$\mathcal C$-term corrections and without any corrections are 
given by
\begin{eqnarray}
\pm \Omega^{(\nicefrac{3}{2})}_{\mathcal C} &=&
       \pm 100.568846(60) \cdot 10^{9} \: {\rm Hz}\, , \label{pred32} \\
\pm \Omega^{(\nicefrac{3}{2})}_{\rm no\,{\mathcal C}} &=&
       \pm  100.568966(27) \cdot 10^{9} \: {\rm Hz}\, , \label{pred32noC}\\
 \pm \sqrt{\Omega^2 + \Delta^2} &=& \pm  99.833975 \cdot 10^{9} \: {\rm Hz}\, .
\end{eqnarray}
Thus also in this case the theoretical 
uncertainties of the generalized corrected 
Rabi frequency in Eq.~(\ref{pred32}) and of the corresponding value
in Eq.~(\ref{pred32noC}) obtained by ignoring the $\mathcal C$-term 
correction shift do not overlap.

The individual shifts with their uncertainties are listed in 
Table~\ref{tabular} 
as discussed in Sec.~\ref{num-1s2p12}.

\section{\label{discuss}DISCUSSION AND SUMMARY}

In this article, we have discussed radiative corrections to the usual 
quantum optical expression for the Mollow spectrum, 
i.e. the resonance fluorescence
spectrum of an atom with two relevant energy levels driven
by a strong coherent laser field. 
To lowest order, the Mollow spectrum consists of one main
peak which is centered at the frequency of the driving laser
field and of two sidebands, which are shifted from the 
center by the generalized Rabi frequency 
$\Omega_{\rm R} = \sqrt{\Omega^2 + \Delta^2}$.
For the analysis, we have used concepts introduced 
originally in two different fields: 
the dressed-state formalism of quantum optics 
and the renormalized radiative corrections which 
are treated in the 
formalism of quantum electrodynamics.
Throughout the analysis, we have adopted the dressed-state basis as the 
natural starting point for our
analysis of corrections to the quasi-energies 
of the combined atom-laser system.

From the point of view of spectroscopy, the Mollow spectrum is 
attractive for several reasons. First of all, the radiative
corrections manifest themselves in the shift of the Mollow
sidebands with respect to the central Mollow peak. 
Thus the tiny radiative corrections are measured relative to 
the generalized Rabi frequency, which for typical parameters
of the driving laser field is several orders of magnitude smaller
than optical frequencies. Also, the Mollow spectrum is centered
around the frequency of the driving laser field. Thus it is
a kind of a differential spectrum because the laser field
detuning is automatically subtracted.

As for quantum electrodynamics, radiative corrections to the Mollow 
spectrum are a quantum-field theoretic problem in the presence of 
two classical background fields, the laser field and the binding 
coulomb field. Our analysis also differs from typical QED
calculations relying on the $S$-matrix formalism, as the process
under study is dynamical. In order to account for the quantum 
fluctuations of the dipole moment leading to the incoherent
part of the Mollow spectrum, a static description is not sufficient.

The shifts of the Mollow sidebands may be interpreted as arising from
corrections to either the detuning or the Rabi frequency.
In particular, the detuning is modified by the bare Lamb shift,
Bloch-Siegert shifts, and virtual off-resonant excitations.
The Rabi frequency is corrected by relativistic and radiative
corrections to the transition dipole moment, by field-configuration
dependent corrections, by a dynamic correction, and by corrections 
to the secular approximation. Of particular interest is 
the dynamical correction to the Rabi frequency. 
This correction arises from an evaluation of the second-order radiative
self-energy corrections of the combined system of atom and laser field
in terms of the dressed states of this system.
To lowest order of the limit $\Omega, \Delta \ll \omega_{eg}$, this
yields a corrections which can be identified with the usual Lamb shift
of the atomic bare states. The dynamic correction is 
then obtained  by keeping terms linear in $\Omega, \Delta$ in the above 
analysis and cannot be explained in terms of the bare state Lamb shift
alone.

The corrections to the detuning may be incorporated into the
analysis by the replacement ($j=\nicefrac{1}{2}, \nicefrac{3}{2}$)
\begin{equation}
\Delta \rightarrow \Delta - \Delta^{(j)}_{\rm rad}\,,
\end{equation}
where $\Delta^{(j)}_{\rm rad}$ is defined in Eq.~(\ref{det-rad}).
Correspondingly, the corrections to the Rabi frequency 
are given by 
\begin{equation}
\Omega \rightarrow \Omega \cdot 
\left (1 + \hat{\Omega}^{(j)}_{\rm rad} \right ).
\end{equation}
The dimensionless quantity $\hat{\Omega}^{(j)}_{\rm rad}$ is defined
in Eq.~(\ref{rabi-rad}). 
Then, the generalized Rabi frequency supplemented by the
discussed relativistic and radiative corrections is 
[Eq.~(\ref{fullOmega})]
\begin{equation}
\Omega^{(j)}_{\mathcal C} =
\sqrt{\Omega^2 \cdot \left(1+{\hat{\Omega}_{\rm rad}^{(j)}}\right)^2 +
\left(\Delta - {\Delta^{(j)}_{\rm rad}}\right)^2}\,.
\end{equation}

In a numerical analysis, we provide a theoretical 
analysis which is required in order
to accurately resolve the dynamical shift.
For this, we suppose the driving laser fields to be in a standing-wave
configuration. As a promising candidate for the experiment, we identify
the hydrogen $1S_{\nicefrac{1}{2}}\leftrightarrow 2P_{\nicefrac{1}{2}}$ and 
$1S_{\nicefrac{1}{2}}\leftrightarrow 2P_{\nicefrac{3}{2}}$ 
transition. The results are discussed 
for a driving laser field parameter set which is expected to be within reach 
of improvements of the currently available Lyman-$\alpha$ laser sources in 
the next few years.

For the $1S_{\nicefrac{1}{2}}\leftrightarrow 2P_{\nicefrac{1}{2}}$ transition
and for $\Omega = 1000\cdot \Gamma_{\nicefrac{1}{2}}$, 
$\Delta = 50 \cdot \Gamma_{\nicefrac{1}{2}}$ ,
the Rabi frequency is shifted
with respect to $\Omega_{\rm R} = \sqrt{\Omega^2 + \Delta^2}$ by
relativistic and radiative corrections as follows,
\begin{equation}
\pm \left (  \Omega^{(\nicefrac{1}{2})}_{\mathcal C} - \Omega_{\rm R} \right ) 
= \pm 738.282(60)\cdot 10^6 \: \rm{Hz}\, .
\end{equation}
The corresponding result for the 
$1S_{\nicefrac{1}{2}}\leftrightarrow 2P_{\nicefrac{3}{2}}$ transition 
with $\Omega = 1000\cdot \Gamma_{\nicefrac{3}{2}}$, $\Delta = 50 \cdot \Gamma_{\nicefrac{3}{2}}$
is
\begin{equation}
\pm \left (  \Omega^{(\nicefrac{3}{2})}_{\mathcal C} - \Omega_{\rm R} \right ) 
= \pm 734.871(60)\cdot 10^6 \: \rm{Hz}\, .
\end{equation}
We note however that we are only concerned with theoretical issues. Thus
uncertainties due to possible experimental issues such as a misalignment
of the apparatus or due to additional trapping potentials have not been
considered. 

In summary, we have presented a detailed analysis
of the leading nonrelativistic and relativistic corrections to
the Mollow spectrum. The analysis includes the relativistic and 
nonrelativistic corrections up to relative orders of $(Z\alpha)^2$ and 
$\alpha(Z\alpha)^2$, respectively, and also includes
Bloch-Siegert shifts, stimulated radiative corrections involving off-resonant
virtual states, field-configuration dependent corrections and
corrections to the secular approximation.
Based on these results, we provide a numerical analysis of the
corrections of the Mollow spectrum of the hydrogen $1S-2P$ transition.
By a comparison with experimental data,
one may verify the presence of dynamical leading-logarithmic correction
to the dressed-state radiative shift, which cannot
be explained in terms of the bare Lamb shift
(see Sec.~\ref{c-term}). This allows to address
questions related to the physical reality of the dressed states.
On the other hand, the comparison with experimental results could
also be used
to interpret the nature of the evaluated radiative corrections in the sense
of the summation formulas which lead to the interpretation of
the shifts as arising from relativistic and radiative corrections
to the detuning and the Rabi frequency.

\section*{Acknowledgments}
Financial support by the German Science Foundation (SFB 276  \& KE 721/1-1)
is gratefully acknowledged. J.~E.~was supported by the German National 
Academic Foundation.
U.~D.~J.~acknowledges helpful conversations with Holger Gies and
Wilhelm Becker regarding the choice of gauge in the description of
dynamical processes.

\appendix

\section{\label{app-dipole} DIPOLE MOMENTS AND SPIN}

The spontaneous emission decay rate of the population
of an excited state $|i\rangle$ to a final state
$|f\rangle$ is given by
\begin{equation}
\Gamma \propto \omega^2 | \langle f |\: \bm{x}\: | i \rangle |^2\, ,
\end{equation}
where all elements of the position vector have to be considered
in the coupling with the vacuum field.
For the $2P \to 1S$ decay in atomic hydrogen,
and in the nonrelativistic Schr\"odinger theory without spin,
the squared modulus of the dipole moment vector is given
by
\begin{equation}
\sum_{i=1}^3 | \langle 1S | \:  x^i\: | 2P \rangle|^2 =
\frac{2^{15}}{3^{10}} \, \frac{1}{(Z\alpha)^2 m^2}\,,
\end{equation}
independent of the magnetic quantum number $m_i~\in~\{-1,0,1\}$
of the ``initial'' $P$ state (by the term ``final state''
we will denote in this section the particular state that enters
as a ``bra-'' in the Dirac notation, i.e.~the 1$S$ state
in the above case). In the Schr\"odinger-Pauli
theory with spin,
one has different decay channels depending on the spin state
of the initial and the final state.
For the channel where the initial and the final state have the same
magnetic quantum number $m_i=m_f=\pm \nicefrac{1}{2}$, one obtains
\begin{equation}
\sum_{i=1}^3 \, | \langle 1S_{\nicefrac{1}{2}}, m_i |\: x^i\: | 
2P_{\nicefrac{1}{2}}, m_i \rangle|^2 =
\frac{1}{3}\, \frac{2^{15}}{3^{10}} \, \frac{1}{(Z\alpha)^2 m^2}\,.
\end{equation}
In contrast, the channel with opposite magnetic quantum number 
$m_i= - m_f = \pm \nicefrac{1}{2}$
yields a contribution of
\begin{equation}
\sum_{i=1}^3 \, | \langle 1S_{\nicefrac{1}{2}}, -m_i |\: x^i\: | 
2P_{\nicefrac{1}{2}}, m_i \rangle|^2 =
\frac{2}{3}\, \frac{2^{15}}{3^{10}} \, \frac{1}{(Z\alpha)^2 m^2}\,.
\end{equation}
This calculation predicts that if one were to measure the electron 
spin polarization in the final state, then the $| 2P_{\nicefrac{1}{2}}\rangle$ state would 
be twice as likely to decay into a $| 1S_{\nicefrac{1}{2}}\rangle$ state with 
opposite total electron angular momentum than into a $| 1S_{\nicefrac{1}{2}}\rangle$ state with the same 
total electron angular momentum as the initial $| 2P_{\nicefrac{1}{2}}\rangle$ state.
Adding the two decay channels to the final state,
we obtain 
\begin{eqnarray}
\label{spinsum}
\sum_{m_f=\pm\nicefrac{1}{2}} &&
\sum_{i=1}^3 \, | \langle 1S_{\nicefrac{1}{2}}, m_f | \: x^i \: | 2P_{\nicefrac{1}{2}}, m_i \rangle|^2  \nonumber \\
&&= \frac{2^{15}}{3^{10}} \, \frac{1}{(Z\alpha)^2 m^2}\,,
\end{eqnarray}
i.e. the same result as in the spinless case, as it should be.
For the decay of the $|2P_{\nicefrac{3}{2}}, m_i\!=\!\pm \nicefrac{1}{2}\rangle$ state, one obtains the same
total decay rate, but here the decay with $m_f=m_i$ is twice as likely
as the decay with $m_f=-m_i$.

These results for the dipole moments  have to be reconsidered for excitation 
of an atom in the $|1S_{\nicefrac{1}{2}}, m_i\rangle$ ground state with a laser field which is linearly polarized in one direction, say 
the $z$-direction. Then other than for the interaction with the vacuum field,
only the $z$-component of the dipole moment vector has to be considered. These matrix
elements can be used to calculate the Rabi flopping frequency corresponding to the driving laser field.
For the spinless case, one obtains
\begin{equation}
\langle 1S | \: z \: | 2P \rangle =
\frac{2^7}{3^5} \, \sqrt{2} \, \frac{1}{Z\alpha m}\,. \label{spinless-z}
\end{equation}
We now include the spin and choose a definite 
initial state $| 1S_{\nicefrac{1}{2}}, m_i=+\nicefrac{1}{2} \rangle$. We obtain
\begin{equation}
\langle 1S_{\nicefrac{1}{2}}, \nicefrac{1}{2} | \: z \: | 2P_{\nicefrac{1}{2}}, \nicefrac{1}{2} \rangle  =
-  \frac{2^7}{3^5} \, \sqrt{\frac{2}{3}} \, \frac{1}{Z\alpha m}\,,
\end{equation}
where we have omitted the ``$m_i=$'' (``$m_f=$'') from the 
initial (final) state vector.
Furthermore, one has
\begin{eqnarray}
&&\langle 1S_{\nicefrac{1}{2}}, \nicefrac{1}{2} |\: z\: | 2P_{\nicefrac{1}{2}}, - \nicefrac{1}{2} \rangle  = 0 \, , \\*
&&\langle 1S_{\nicefrac{1}{2}}, \nicefrac{1}{2} |\: x\: | 2P_{\nicefrac{1}{2}}, \nicefrac{1}{2} \rangle  = 0 \, , \\*
&&\langle 1S_{\nicefrac{1}{2}}, \nicefrac{1}{2} |\: y\: | 2P_{\nicefrac{1}{2}}, \nicefrac{1}{2} \rangle  = 0 \, , 
\end{eqnarray}
so that the $z$-polarized field only couples 
the $2P_{\nicefrac{1}{2}}$ state with $m_i=m_f$ to 
the ground state, and this excited state is only one coupled 
to the ground state by $z$-polarized light.
For the $|2P_{\nicefrac{3}{2}},m_f\rangle$ upper state, the corresponding results are
\begin{align}
&\langle 1S_{\nicefrac{1}{2}}, \nicefrac{1}{2} |\: z\: | 2P_{\nicefrac{3}{2}}, \nicefrac{1}{2} \rangle  =
 \frac{2^8}{3^5} \, \sqrt{\frac{1}{3}} \, \frac{1}{Z\alpha m} \, , \\
&\langle 1S_{\nicefrac{1}{2}}, \nicefrac{1}{2} |\: z\: | 2P_{\nicefrac{3}{2}}, -\nicefrac{1}{2} \rangle  = 0 \, , \\
&\langle 1S_{\nicefrac{1}{2}}, \nicefrac{1}{2} |\: x\: | 2P_{\nicefrac{3}{2}}, \nicefrac{1}{2} \rangle  = 0 \, , \\
&\langle 1S_{\nicefrac{1}{2}}, \nicefrac{1}{2} |\: y\: | 2P_{\nicefrac{3}{2}}, \nicefrac{1}{2} \rangle  = 0 \, , \\
&\langle 1S_{\nicefrac{1}{2}}, \nicefrac{1}{2} |\: z\: | 2P_{\nicefrac{3}{2}}, \pm \nicefrac{3}{2} \rangle  = 0 \, .
\end{align}
Thus also in this case only the upper state with same magnetic quantum number
is coupled to the ground state, but with a matrix element which differs
by a factor of $-\sqrt{2}$ in magnitude from the corresponding result for the
$|2P_{\nicefrac{1}{2}},\nicefrac{1}{2}\rangle$ upper state.
From these spin-resolved results, the corresponding matrix element
without spin Eq.~(\ref{spinless-z}) may be obtained by summing over the 
final states and averaging over the initial states.

\section{\label{app-derivation}EVALUATION OF THE MATRIX ELEMENTS}

In this section we demonstrate the evaluation of the
matrix element in Eq. (\ref{m1}):
\begin{equation}
M_e = \left< e \left| \, z\,  \,
G^{''}(\zeta) \, z\,  \right| e \right> \, ,
\end{equation}
where $|e\rangle$ is the 2$P$, $m=0$ state.
We start by calculating the ``unreduced'' matrix element
\begin{equation}
\overline{M}_e(\zeta) = \left< e \left| \, z\, 
G(\zeta) \, z\,  \right| e \right> 
\end{equation}
where the full sum over intermediate states is employed in
$G(\zeta)$, and the
wavefunction of the excited state state 
$\Phi_{2P,m=0}$ is given by a product of a radial 
and an angular contribution:
\[
\Phi_{2P,m=0}(\vec{r}) = R_{2P}(r) \: Y_{10}(\theta, \phi) \, .
\]
Here, $(r,\theta,\phi)$ are spherical coordinates.
In these coordinates a representation of the Green 
function in position space is given by~\cite[Eq. 2.2]{SwDr1991b}
\[
\frac{1}{H-E} = \sum _{l, m} g_l (r_1, r_2, \nu) \: 
Y_{lm}(\theta_1, \phi_1)\: Y^*_{lm}(\theta_2, \phi_2)
\]
where $g_n(r_1, r_2, \nu)$ is the radial component of the Schr\"odinger-Coulomb propagator 
\begin{eqnarray}
g_l(r_1, r_2, \nu) &=& 2m \left ( \frac{2}{a_{\rm B}\nu} \right )^{2l+1} 
(r_1r_2)^l \exp \left ( - \frac{r_1+r_2}{a_{\rm B}\nu} \right ) \nonumber \\
&& \times \sum _{k=0}^{\infty} \frac{k! \: L_k^{2l+1} 
\left ( \frac{2r_1}{a_{\rm B}\nu} \right )   \: 
L_k^{2l+1} \left ( \frac{2r_2}{a_{\rm B}\nu} \right )}{(2l+1+k)!\:(l+1+k-\nu)}
\end{eqnarray}
containing associated Laguerre polynomials $L^b_a(r)$.
The quantity $\nu~=~(a_{\rm B}\sqrt{-2mE})^{-1}$ is an energy
parameter which is related to the parameter $t$ used in Sec.~\ref{sec-offres}
by $\nu~=~nt$ where $n$ is the principal quantum number of the initial bound
state. The Bohr radius is defined in (\ref{aBohr}), 
and we evaluate all matrix elements here for the case $Z=1$
(atomic hydrogen).
Thus for the 2$P$ state discussed here we have $\nu~=~2t$.
The index $l$ is summed over all possible angular momentum numbers
of the virtual intermediate states. Starting from a $P$ state with $l=1$, both
$S$ ($l=0$) and $D$ ($l=2$) states are possible as intermediate states.
The integration may be further separated in angular and radial parts:
\begin{equation}
\overline{M}_e = 
\overline{M}_{l=0}^{\rm ang} \cdot \overline{M}_{l=0}^{\rm rad}+
\overline{M}_{l=2}^{\rm ang} \cdot \overline{M}_{l=2}^{\rm rad}\,,
\label{melem}
\end{equation}
where the angular integrations yield
\begin{eqnarray}
\overline{M}_{l=0}^{\rm ang} = \frac{1}{3}\, , \qquad 
\overline{M}_{l=2}^{\rm ang} = \frac{4}{15}\, .
\end{eqnarray}
The radial parts may written explicitly as
\begin{eqnarray}
\overline{M}_{l}^{\rm rad}&=& \int_0^\infty 
dr_1 \, dr_2 \, r_1^3 \, r_2^3 \, 
R_{2P}^*(r_1) \, R_{2P}(r_2) \nonumber \\
&& \qquad \qquad \times \: g_l(r_1,r_2,\nu)  \, ,
\end{eqnarray}
for $l=0,2$.
Simplifying further, one obtains 
\begin{eqnarray}
\overline{M}_{l=0}^{\rm rad}&=&  \frac{m}{12 t a_{\rm B}^6}\, 
\sum_{k=0}^{\infty} \frac{k!}{(1+k)!(1+k-2t)} \, I_0^2\, , \\
I_0&=& \int_{0}^{\infty} dr \, r^4 \,
L_k^{1} \left ( \frac{r}{a_{\rm B} t} \right ) \,
e^{-\frac{1+t}{2 a_{\rm B} t} r} \, .
\end{eqnarray}

The integral in $I_0$ can be evaluated using
(see~\cite[Sec.~6.10]{Ba1953vol1} and~\cite[Sec.~10.12]{Ba1953vol2})
\begin{eqnarray}
&&\int _0 ^\infty {\rm d}r \: \exp (-\lambda r) \: r^\gamma\: L_n^\mu(r) = 
\frac{\lambda^{-1-\gamma}\: \Gamma (\gamma+1)}{n!\: \Gamma (\mu+1)} 
\times \nonumber \\*
&&\times \:  \Gamma (\mu+n+1) \: _2F_1 (-n, \gamma+1, \mu+1, \lambda^{-1})  
\end{eqnarray}
to give
\begin{equation}
I_0 =  \frac{768\, a_{\rm B}^5 \,t^5}{(1+t)^5} \: 
\frac {\Gamma(k+2)}{k!} \: 
{}_2F_1 \left (-k, 5, 2; \frac{2}{1+t}\right )\, .
\end{equation}
Using an explicit expression for the hypergeometric 
function~\cite[Sec.~2.1.1]{Ba1953vol1},
we obtain
\begin{equation}
{}_2F_1 (a,b,c;z) = \sum _{j=0}^{\infty} 
\frac{(a)_j (b)_j}{(c)_j}\: \frac{z^j}{j!}\, , 
\end{equation}
where the Pochhammer symbols $(a)_j$ are given by
\begin{equation}
(a)_j = \frac{\Gamma (a+j)}{\Gamma (a)} \, .
\end{equation}
Contiguous relations for the hypergeometric 
function~\cite[Sec.~2.8]{Ba1953vol1}
then lead to
\begin{eqnarray*}
\overline{M}^{\rm rad}_{l=0}&=&m a_{\rm B}^4
\left ( \frac{16 t^2 {\cal X}_0(t)}{3(t-1)^6(t+1)^4} 
- \frac{2^{14} t^{11}}{3(t^2-1)^6} \Phi(2,t) \right ) ,\\
{\cal X}_0(t) &=& 45 - 90t - 84t^2 + 258t^3 + 18t^4 - 294t^5 \\
&&+ 148t^6 - 2t^7 + 257t^8 \, ,
\end{eqnarray*}
where the hypergeometric function $\Phi(n,t)$ is defined in
Sec.~\ref{sec-offres}.
A similar calculation for the intermediate $D$ ($l=2$) states yields
\begin{eqnarray*}
\overline{M}^{\rm rad}_{l=2}&=&
m a_{\rm B}^4\: 
\left ( \frac{16\, t^2\, {\cal X}_2(t)}{3(t-1)^7(t+1)^5} \right . \\
&&\left . \qquad 
- \frac{2^{16}\,t^{11}\,(4t^2-1)}{3(t^2-1)^7} \Phi(2,t)\, \right ) \, ,\\
{\cal X}_2(t) &=& -45 + 90t + 165t^2 - 420t^3 - 174t^4 + 768t^5 \\
&& - 34t^6 - 700t^7 - 37t^8 - 1274t^9 + 4733t^{10} \, .
\end{eqnarray*}
Inserting in Eq.~(\ref{melem}) finally yields 
the expression in Eq. (\ref{matrix2p}).
The reduced matrix element can then be obtained from this
by subtracting the contributions of the two intermediate states 
$|e\rangle, |g\rangle$.
The excited state contribution vanishes due to parity, and the
ground state contribution is given by 
\begin{equation}
\frac{\left | \left< g \left| z \right | e \right > \right |^2}
{E_{1S} - \zeta}\, . \label{reduktion}
\end{equation}
This term, which cancels the divergence as $\zeta \to E_{1S}$
in (\ref{cancellation}), may be verified by
inserting the resonant 1$S$ state as the intermediate
state into the matrix element. Alternatively,
the cancellation may be seen as follows:
On setting the intermediate
state energy to $E_{1S} + \epsilon$, the series
expansion of the unreduced matrix 
element $\overline{M}_e(E_{1S})$ receives a
contribution proportional to $1/\epsilon$ which diverges
for $\epsilon \to 0$. This diverging part is canceled
by the intermediate state contribution Eq. (\ref{reduktion})
in the reduced matrix element $M_e(\zeta)$ to give a finite
result.

If one compares this derivation with a similar calculation
for the standard matrix element
\begin{eqnarray}
\overline{\cal M} &=& \sum_{i=1}^{3} \left< e \left| x^i \,
G^{''}(\zeta) \, x^i \right| e \right>\, , \label{appendix-std}\\
&=& \overline{\cal M}_{l=0}^{\rm ang} \cdot \overline{\cal M}_{l=0}^{\rm rad}+
 \overline{\cal M}_{l=2}^{\rm ang} \cdot \overline{\cal M}_{l=2}^{\rm rad} \, ,
\end{eqnarray}
where all polarization directions are considered,
one finds that the respective radial parts for $l=0$ and $l=2$
are identical to the ones in Eq.~(\ref{melem}):
\begin{equation}
\overline{\cal M}_{l=0}^{\rm rad} = \overline{M}_{l=0}^{\rm rad} \, ,\qquad
\overline{\cal M}_{l=2}^{\rm rad} = \overline{M}_{l=2}^{\rm rad} \, .
\end{equation}
For the angular parts however one finds
\begin{equation}
\overline{\cal M}_{l=0}^{\rm ang} = 
\overline{M}_{l=0}^{\rm ang}  \, ,\qquad
\overline{\cal M}_{l=2}^{\rm ang} = 
\frac{5}{2}\: \overline{M}_{l=2}^{\rm ang} \, .
\end{equation}
The reason for this is that the fixed polarization of the driving laser
field only allows to excite one of the magnetic sublevels of the 
intermediate $S$ and $D$ states. The sum over all polarizations in the
standard matrix element still only gives one magnetic sublevel for
the intermediate $S$ states, but three possible virtual $D$ states.
Due to this asymmetry the desired matrix element Eq.~(\ref{matrix2p}) cannot
be calculated directly from the standard matrix element in
Eq.~(\ref{appendix-std}).


\begin{thebibliography}{10}

\bibitem{ReEtAl2000}
J. Reichert, M. Niering, R. Holzwarth, M. Weitz, T. Udem, and T.~W. H\"{a}nsch,
  Phys. Rev. Lett. {\bf 84},  3232  (2000).

\bibitem{MoTa2000}
P.~J. Mohr and B.~N. Taylor, Rev. Mod. Phys. {\bf 72},  351  (2000).

\bibitem{PrTjMa1995}
J.~D. Prestage, R.~J. Tjoelker, and L. Maleki, Phys. Rev. Lett. {\bf 74},  3511
   (1995).

\bibitem{WeFlChDrBa1999}
J.~K. Webb, V.~V. Flambaum, C.~W. Churchill, M.~J. Drinkwater, and J.~D.
  Barrow, Phys. Rev. Lett. {\bf 82},  884  (1999).

\bibitem{DzFlWe1999}
V.~A. Dzuba, V.~V. Flambaum, and J.~K. Webb, Phys. Rev. Lett. {\bf 82},  888
  (1999).

\bibitem{WeEtAl2003}
J.~K. Webb, M.~T. Murphy, V.~V. Flambaum, V.~A. Dzuba, J.~D. Barrow, C.~W.
  Churchill, J.~X. Prochaska, and A.~M. Wolfe, Phys. Rev. Lett. {\bf 87},
  091301  (2001).

\bibitem{MuWeFl2003}
M.~T. Murphy, J.~K. Webb, and V.~V. Flambaum, Mon. Not. Roy. Astron. Soc. {\bf
  345},  609  (2003).

\bibitem{Uz2003}
J.-P. Uzan, Rev. Mod. Phys. {\bf 75},  403  (2003).

\bibitem{MaEtAl2003}
H. Marion, F.~P.~D. Santos, M. Abgrall, S. Zhang, Y. Sortais, S. Bize, I.
  Maksimovic, D. Calonico, S. Bize, J. Gr\"unert, C. Mandache, P. Lemonde,
  G.~S.~P. Laurent, A. Clairon, and C. Salomon, Phys. Rev. Lett. {\bf 90},
  150801  (2003).

\bibitem{FiEtAl2004}
M. Fischer, N. Kolachevsky, M. Zimmermann, R. Holzwarth, T. Udem, T.~W.
  H\"{a}nsch, M. Abgrall, J. Gr\"unert, I. Maksimovic, S. Bize, H. Marion,
  F.~P.~D. Santos, P. Lemonde, G. Santarelli, A. Clairon, C. Salomon, M. Haas,
  U.~D. Jentschura, and C.~H. Keitel, Phys. Rev. Lett. {\bf 92},  230802
  (2004).

\bibitem{Pa1991}
K. Pachucki, Phys. Rev. A {\bf 44},  5407  (1991).

\bibitem{LaRe1950}
W.~E. Lamb and R.~C. Retherford, Phys. Rev. {\bf 79},  549  (1950).

\bibitem{La1952}
W.~E. Lamb, Phys. Rev. {\bf 85},  259  (1952).

\bibitem{PoZi1959}
E.~A. Power and S. Zienau, Phil. Trans. Roy. Soc. London A {\bf 251},  427
  (1959).

\bibitem{Ya1976}
K.~H. Yang, Ann. Phys. (N. Y.) {\bf 101},  62  (1976).

\bibitem{FoQuBa1977}
J.~J. Forney, A. Quattropani, and F. Bassani, Nuovo Cim. B {\bf 37},  78
  (1977).

\bibitem{Ko1978prl}
D.~H. Kobe, Phys. Rev. Lett. {\bf 40},  538  (1978).

\bibitem{BrScScZuGo1983}
W. Becker, R.~R. Schlicher, M.~O. Scully, M.~S. Zubairy, and M. Goldhaber,
  Phys. Lett. B {\bf 131},  16  (1983).

\bibitem{BeScSc1984}
W. Becker, R.~R. Schlicher, and M.~O. Scully, Phys. Lett. A {\bf 106},  441
  (1984).

\bibitem{ScBeBeSc1984}
R.~R. Schlicher, W. Becker, J. Bergou, and M.~O. Scully, in {\em Quantum
  Electrodynamics and Quantum Optics} (A. O. Barut, Ed.), Plenum (New York),
  pp.~405-441 (1984).

\bibitem{LaScSc1987}
W.~E. Lamb, R.~R. Schlicher, and M.~O. Scully, Phys. Rev. A {\bf 36},  2763
  (1987).

\bibitem{ScZu1997}
M.~O. Scully and M.~S. Zubairy, {\em Quantum Optics} (Cambridge University
  Press, Cambridge, 1997).

\bibitem{JaCu1963}
E.~T. Jaynes and F.~W. Cummings, Proc. IEEE {\bf 51},  89  (1963).

\bibitem{Mo1969}
B.~R. Mollow, Phys. Rev. {\bf 188},  1969  (1969).

\bibitem{CT1975misc}
C. Cohen-Tannoudji, {\em Atoms in Strong Resonant Fields}, in {\em Aux
  fronti\`{e}res de la spectroscopie laser/Frontiers in Laser Spectroscopy},
  Eds.~R. Balian, S. Haroche and S. Liberman, Eds., pp.~4--104 (North--Holland,
  Amsterdam, 1975).

\bibitem{AuTo1955}
S.~H. Autler and C.~H. Townes, Phys. Rev. {\bf 100},  703  (1955).

\bibitem{Kr1982}
G.~Y. Kryuchkov, JETP {\bf 56},  1153  (1982), [Zh. \'{E}ksp. Teor. Fiz. {\bf
  83}, 1992 (1982)].

\bibitem{JeEvHaKe2003}
U.~D. Jentschura, J. Evers, M. Haas, and C.~H. Keitel, Phys. Rev. Lett. {\bf
  91},  253601  (2003).

\bibitem{JeKe2004}
U.~D. Jentschura and C.~H. Keitel, Ann. Phys. (N. Y.) {\bf 310},  1  (2004).

\bibitem{EiWaHa2001}
K.~S.~E. Eikema, J. Walz, and T.~W. H\"{a}nsch, Phys. Rev. Lett. {\bf 86},
  5679  (2001).

\bibitem{Pa2002}
A. Pahl, {\em PhD thesis: Erzeugung von kontinuierlicher koh\"arenter
  Lyman-$\alpha$-Strahlung zur 1S-2P-Spektroskopie an Antiwasserstoff (in
  German)} (University of Munich, 2002, unpublished).

\bibitem{CTDRGr1992}
C. Cohen-Tannoudji, J. Dupont-Roc, and G. Grynberg, {\em Atom--Photon
  Interactions} (J. Wiley \& Sons, New York, 1992).

\bibitem{ItZu1980}
C. Itzykson and J.~B. Zuber, {\em Quantum Field Theory} (McGraw-Hill, New York,
  NY, 1980).

\bibitem{JePa1996}
U.~D. Jentschura and K. Pachucki, Phys. Rev. A {\bf 54},  1853  (1996).

\bibitem{PaJe2003}
K. Pachucki and U.~D. Jentschura, Phys. Rev. Lett. {\bf 91},  113005  (2003).

\bibitem{KrHe1925}
W. Kramers and W.~H. Heisenberg, Z. Phys. {\bf 31},  681  (1925).

\bibitem{SaYe1990}
J. Sapirstein and D.~R. Yennie, in {\em Quantum Electrodynamics} (T. Kinoshita,
  Ed.), World Scientific (Singapore), pp.~560-672 (1990).

\bibitem{BlSi1940}
F. Bloch and A.~J. Siegert, Phys. Rev. {\bf 57},  522  (1940).

\bibitem{BrKe2000}
D.~E. Browne and C.~H. Keitel, J. Mod. Opt. {\bf 47},  1307  (2000).

\bibitem{Pa1993}
K. Pachucki, Ann. Phys. (N. Y.) {\bf 226},  1  (1993).

\bibitem{JeSoMo1997}
U.~D. Jentschura, G. Soff, and P.~J. Mohr, Phys. Rev. A {\bf 56},  1739
  (1997).

\bibitem{ShBe1990}
N. Shafer and R. Bersohn, Phys. Rev. A {\bf 42},  1313  (1990).

\bibitem{Ya2003}
V. Yakhontov, Phys. Rev. Lett. {\bf 91},  093001  (2003).

\bibitem{SwDr1991a}
R.~A. Swainson and G.~W.~F. Drake, J. Phys. A {\bf 24},  79  (1991).

\bibitem{SwDr1991b}
R.~A. Swainson and G.~W.~F. Drake, J. Phys. A {\bf 24},  95  (1991).

\bibitem{Je1996}
U.~D. Jentschura, {\em Master Thesis: The Lamb Shift in Hydrogenlike Systems
  [in German: Theorie der Lamb--Verschiebung in wasserstoffartigen Systemen]}
  (University of Munich, 1996, unpublished, available as e-print
  hep-ph/0306065).

\bibitem{Pa2004}
K. Pachucki, Phys. Rev. A {\bf 69},  052502  (2004).

\bibitem{Ka1996}
S.~G. Karshenboim, J. Phys. B {\bf 29},  L29  (1996).

\bibitem{IvKa1996}
V.~G. Ivanov and S.~G. Karshenboim, Phys. Lett. A {\bf 210},  313  (1996).

\bibitem{SaPaCh2004}
J. Sapirstein, K. Pachucki, and K.~T. Cheng, Phys. Rev. A {\bf 69},  022113
  (2004).

\bibitem{Ba1953vol1}
H. Bateman, {\em Higher Transcendental Functions} (McGraw-Hill, New York, NY,
  1953), Vol.~1.

\bibitem{Ba1953vol2}
H. Bateman, {\em Higher Transcendental Functions} (McGraw-Hill, New York, NY,
  1953), Vol.~2.

\end{thebibliography}
\end{document}